\documentclass[a4paper,11pt]{article}
\pdfoutput=1 

\usepackage{jheppub} 

\usepackage[T1]{fontenc} 
\usepackage{latexsym}
\usepackage{amsmath,amssymb}
\usepackage{array}
\usepackage{float}
\newcommand{\lc}{\varepsilon}

\newcommand{\ds}{\displaystyle}
\newcommand{\del}{\partial}
\newcommand{\itGamma}{{\mit{\Gamma}}}
\newcommand{\D}{\mathrm{d}}

\newcommand{\diag}{\mathop{\rm diag}\nolimits}
\newcommand{\rmd}{\mathrm{d}}
\usepackage{xcoffins}
\NewCoffin\tablecoffin
\NewDocumentCommand\Vcentre{m}
  {%
    \SetHorizontalCoffin\tablecoffin{#1}%
    \TypesetCoffin\tablecoffin[l,vc]%
  }
\newcommand{\killing}[2]{ {\langle #1 , #2 \rangle } }

\newcommand{\realni}{\ensuremath{\mathbb{R}}}

\newcommand{\grasmanovi}{\ensuremath{\mathbb{G}}}

\newcommand{\cA}{{\cal A}}
\newcommand{\cD}{{\cal D}}
\newcommand{\cF}{{\cal F}}
\newcommand{\cG}{{\cal G}}
\newcommand{\cH}{{\cal H}}

\newcommand{\cM}{{\cal M}}

\newcommand{\cT}{{\cal T}}
\newcommand{\cS}{{\cal S}}
\newcommand{\cX}{{\cal X}}
\newtheorem{Theorem}{Theorem}

\newtheorem{Definition}[Theorem]{Definition}

\newtheorem{Fundamental Theorem}{Fundamental Theorem}

\title{Higher Gauge Theories Based on 3-groups}

\author[1]{T. Radenkovi\' c\note{Corresponding author.}}
\author{and M. Vojinovi\' c}

\affiliation{Institute of Physics, University of Belgrade, \\ Pregrevica 118, 11080 Belgrade, Serbia}

\emailAdd{rtijana@ipb.ac.rs}
\emailAdd{vmarko@ipb.ac.rs}

\abstract{We study the categorical generalizations of a $BF$ theory to $2BF$ and $3BF$ theories, corresponding to $2$-groups and $3$-groups, in the framework of higher gauge theory. In particular, we construct the constrained $3BF$ actions describing the correct dynamics of Yang-Mills, Klein-Gordon, Dirac, Weyl, and Majorana fields coupled to Einstein-Cartan gravity. The action is naturally split into a topological sector and a sector with simplicity constraints, adapted to the spinfoam quantization programme. In addition, the structure of the $3$-group gives rise to a novel gauge group which specifies the spectrum of matter fields present in the theory, just like the ordinary gauge group specifies the spectrum of gauge bosons in the Yang-Mills theory. This allows us to rewrite the whole Standard Model coupled to gravity as a constrained $3BF$ action, facilitating the nonperturbative quantization of both gravity and matter fields. Moreover, the presence and the properties of this new gauge group open up a possibility of a nontrivial unification of all fields and a possible explanation of fermion families and all other structure in the matter spectrum of the theory.}


\arxivnumber{1904.07566}

\begin{document} 
\maketitle
\flushbottom

\section{\label{SecIntroduction}Introduction}

The quantization of the gravitational field is one of the most prominent open problems in modern theoretical physics. Within the Loop Quantum Gravity framework, one can study the nonperturbative quantization of gravity, both canonically and covariantly, see \cite{RovelliBook, RovelliVidottoBook, Thiemann2007} for an overview and a comprehensive introduction. The covariant approach focuses on the definition of the path integral for the gravitational field,
\begin{equation}
Z = \int \cD g\; e^{iS[g]}\,,
\end{equation}
by considering a triangulation of a spacetime manifold, and defining the path integral as a discrete state sum of the gravitational field configurations living on the simplices in the triangulation. This quantization technique is known as the {\em spinfoam} quantization method, and roughly goes along the following lines:
\begin{enumerate}
\item first, one writes the classical action $S[g]$ as a topological $BF$ action plus a simplicity constraint,
\item then one uses the algebraic structure (a Lie group) underlying the topological sector of the action to define a triangulation-independent state sum $Z$,
\item and finally, one imposes the simplicity constraints on the state sum, promoting it into a path integral for a physical theory.
\end{enumerate}
This quantization prescription has been implemented for various choices of the action, the Lie group, and the spacetime dimension. For example, in $3$ dimensions, the prototype spinfoam model is known as the Ponzano-Regge model \cite{PonzanoRegge1968}. In $4$ dimensions there are multiple models, such as the Barrett-Crane model \cite{BarrettCrane,BarrettCrane1}, the Ooguri model \cite{Ooguri}, and the most sophisticated EPRL/FK model \cite{EPRL,FK}. All these models aim to define a viable theory of quantum gravity, with variable success. However, virtually all of them are focused on pure gravity, without matter fields. The attempts to include matter fields have had limited success \cite{RovelliSpinfoamFermions}, mainly because the mass terms could not be expressed in the theory due to the absence of the tetrad fields from the $BF$ sector of the theory.

In order to resolve this issue, a new approach has been developed, using the categorical generalization of the $BF$ action, within the framework of {\em higher gauge theory} (see \cite{BaezHuerta2011} for a review). In particular, one uses the idea of a categorical ladder to promote the $BF$ action, which is based on some Lie group, into a $2BF$ action, which is based on the so-called $2$-group structure. If chosen in a suitable way, the $2$-group structure should hopefully introduce the tetrad fields into the action. This approach has been successfully implemented \cite{MikovicVojinovic2012}, rewriting the action for general relativity as a constrained $2BF$ action, such that the tetrad fields are present in the topological sector. This result opened up a possibility to couple all matter fields to gravity in a straightforward way. Nevertheless, the matter fields could not be naturally expressed using the underlying algebraic structure of a $2$-group, rendering the spinfoam quantization method only half-implementable, since the matter sector of the classical action could not be expressed as a topological term plus a simplicity constraint, which means that the steps 2 and 3 above could not be performed for the matter sector of the action.

We address this problem in this paper. As we will show, it turns out that it is necessary to perform one more step in the categorical ladder, generalizing the underlying algebraic structure from a $2$-group to a $3$-group. This generalization then naturally gives rise to the so-called $3BF$ action, which proves to be suitable for a unified description of both gravity and matter fields. The steps of the categorical ladder can be conveniently summarized in the following table:

\begin{center}
\setlength{\tabcolsep}{0pt}
     \label{tab:table5}\begin{tabular}{|c|c|c|c|c|} \hline
\shortstack{categorical$\vphantom{A^A}$ \\ structure} & \shortstack{algebraic$\vphantom{A^A}$ \\ structure} & \shortstack{linear$\vphantom{gA^A}$ \\ structure} & \shortstack{topological$\vphantom{A^A}$ \\ action} & \shortstack{degrees of$\vphantom{A^A}$ \\ freedom} \\ \hline\hline
Lie group$\vphantom{\ds\int}$ & Lie group & Lie algebra & $BF$ theory & gauge fields \\ \hline
Lie $2$-group$\vphantom{\ds\int}$ & \raisebox{-0.5em}[1em][0em]{\shortstack{Lie crossed\\ module}} & \raisebox{-0.5em}[1em][0em]{\shortstack{differential Lie \\ crossed module}} & $2BF$ theory & tetrad fields \\ \hline
\ \ Lie $3$-group$\vphantom{\ds\int}$\ \ \  &\ \  \raisebox{-0.5em}[1em][0em]{\shortstack{Lie $2$-crossed\\ module}}\ \ \  &\ \  \raisebox{-0.5em}[1em][0em]{\shortstack{differential Lie \\ $2$-crossed module}}\ \ \  &\ \  $3BF$ theory\ \ \  & \ \ \raisebox{-0.5em}[1em][0em]{\shortstack{scalar and \\ fermion fields}}\ \ \  \\ \hline
\end{tabular}
\end{center}

Once the suitable gauge $3$-group has been specified and the corresponding $3BF$ action constructed, the most important thing that remains, in order to complete the step 1 of the spinfoam quantization programme, is to impose appropriate simplicity constraints onto the degrees of freedom present in the $3BF$ action, so that we obtain the desired classical dynamics of the gravitational and matter fields. Then one can proceed with steps 2 and 3 of the spinfoam quantization, hopefully ending up with a viable model of quantum gravity and matter.

In this paper, we restrict our attention to the first of the above steps: we will construct a constrained $3BF$ action for the cases of Klein-Gordon, Dirac, Weyl and Majorana fields, as well as Yang-Mills and Proca vector fields, all coupled to the Einstein-Cartan gravity in the standard way. This construction will lead us to an unexpected novel result. As we shall see, the scalar and fermion fields will be {\em naturally associated to a new gauge group}, generalizing the notion of a gauge group in the Yang-Mills theory, which describes vector bosons. This new group opens up a possibility to use it as an algebraic way of classifying matter fields, describing the structures such as quark and lepton families, and so on. The insight into the existence of this new gauge group is the consequence of the categorical ladder and is one of the main results of the paper. However, given the complexity of the algebraic properties of $3$-groups, we will restrict ourselves only to the reconstruction of the already known theories, such as the Standard Model (SM), in the new framework. In this sense, any potential explanation of the spectrum of matter fields in the SM will be left for future work.

The layout of the paper is as follows. In subsection \ref{subAII} we will give a short overview of the constrained $BF$ actions, including the well-known example of the Plebanski action for general relativity, and a completely new example of the Yang-Mills theory rewritten as a constrained $BF$ model. In the subsection \ref{subBII} we also introduce the formalism of the constrained $2BF$ actions, reviewing the example of general relativity as a constrained $2BF$ action, first introduced in \cite{MikovicVojinovic2012}. In addition, we will demonstrate how to couple gravity in a natural way within the formalism of $2$-groups. Section \ref{SecIII} contains the main results of the paper and is split into $4$ subsections. The subsection \ref{sec:III.1} introduces the formalism of $3$-groups, and the definition and properties of a $3BF$ action, including the three types of gauge transformations. The subsection \ref{sec:III.2}
focuses on the construction of a constrained $3BF$ action which describes a single real scalar field coupled to gravity. It provides the most elementary example of the insight that matter fields correspond to a gauge group. Encouraged by these results, in the subsection \ref{sec:III.3} we construct the constrained $3BF$ action for the Dirac field coupled to gravity and specify its gauge group. Finally, the subsection \ref{sec:III.4} deals with the construction of the constrained $3BF$ action for the Weyl and Majorana fields coupled to gravity, thereby covering all types of fields potentially relevant for the Standard Model and beyond. After the construction of all building blocks, in section \ref{SecIV} we apply the results of sections \ref{SecII} and \ref{SecIII} to construct the constrained $3BF$ action corresponding to the full Standard Model coupled to Einstein-Cartan gravity. Finally, section \ref{SecV} is devoted to the discussion of the results and the possible future lines of research. The Appendices contain some mathematical reminders and technical details.

The notation and conventions are as follows. The local Lorentz indices are denoted by the Latin letters $a,b,c,\dots$, take values $0,1,2,3$, and are raised and lowered using the Minkowski metric $\eta_{ab}$ with signature $(-,+,+,+)$. Spacetime indices are denoted by the Greek letters $\mu,\nu,\dots$, and are raised and lowered by the spacetime metric $g_{\mu\nu} = \eta_{ab} e^a{}_{\mu} e^b{}_{\nu}$, where $e^a{}_{\mu}$ are the tetrad fields. The inverse tetrad is denoted as $e^{\mu}{}_a$. All other indices that appear in the paper are dependent on the context, and their usage is explicitly defined in the text where they appear. A lot of additional notation is defined in Appendix A. We work in the natural system of units where $c=\hbar=1$, and $G = l_p^2$, where $l_p$ is the Planck length.

\section{\label{SecII}$BF$ and $2BF$ models, ordinary gauge fields and gravity}

Let us begin by giving a short review of $BF$ and $2BF$ theories in general. For additional information on these topics, see for example \cite{BFgravity2016, zakopane, plebanski1977, BaezHuerta2011, GirelliPfeifferPopescu2008, FariaMartinsMikovic2011, crane2003}.

\subsection{$BF$ theory}\label{subAII}

Given a Lie group $G$ and its corresponding Lie algebra $\mathfrak{g}$, one can introduce the so-called $BF$ action as
\begin{equation}\label{eq:bf}
    S_{BF} =\int_{\cM_4} \langle B \wedge \cF \rangle_\mathfrak{g}\,.
\end{equation}
Here, $\cF\equiv \D \alpha+ \alpha\wedge \alpha$ is the curvature $2$-form for the algebra-valued connection $1$-form $\alpha \in \cA^1(\cM_4\,, \mathfrak{g})$ on some $4$-dimensional spacetime manifold $\cM_4$. In addition, $B \in \cA^2(\cM_4\,, \mathfrak{g})$ is a Lagrange multiplier $2$-form, while $\langle \_\,,\_\rangle{}_{\mathfrak{g}}$ denotes the $G$-invariant bilinear symmetric nondegenerate form.

From the structure of (\ref{eq:bf}), one can see that the action is diffeomorphism invariant, and it is usually understood to be gauge invariant with respect to $G$. In addition to these properties, the $BF$ action is topological, in the following sense. Varying the action (\ref{eq:bf}) with respect to $B^{\beta}$ and $\alpha^{\beta}$, where the index $\beta$ counts the generators of $\mathfrak{g}$ (see Appendix \ref{ApendiksA} for notation and conventions), one obtains the equations of motion of the theory,
\begin{equation}
    \cF=0\,,\quad \quad \nabla B \equiv \D B + \alpha \wedge B  =0\,.
\end{equation}
From the first equation of motion, one immediately sees that $\alpha$ is a flat connection, which then together with the second equation of motion implies that $B$ is constant. Therefore, there are no local propagating degrees of freedom in the theory, and one then says that the theory is topological.

Usually, in physics one is interested in theories which are nontopological, i.e., which have local propagating degrees of freedom. In order to transform the $BF$ action into such a theory, one adds an additional term to the action, commonly called the {\em simplicity constraint}. A very nice example is the Yang-Mills theory for the $SU(N)$ group, which can be rewritten as a constrained $BF$ theory in the following way:
\begin{equation}\label{eq:bfgauge}
    S=\int B_I\wedge F^I+ \lambda^I\wedge \Big(B_I-\frac{12}{g}{M_{ab}}_I\delta^a\wedge \delta^b \Big)+\zeta^{ab}{}^I \Big( {M_{ab}}{}_I\varepsilon_{cdef}\delta^c\wedge \delta^d \wedge \delta^e \wedge \delta^f- g_{IJ}F^J \wedge \delta_a \wedge \delta_b \Big) \,.
\end{equation}
Here $F \equiv \D A + A\wedge A$ is again the curvature $2$-form for the connection $A \in \cA^1(\cM_4\,, \mathfrak{su}(N))$, and $B\in \cA^2(\cM_4\,, \mathfrak{su}(N))$ is the Lagrange multiplier $2$-form. The Killing form $g_{IJ} \equiv \killing{\tau_I}{\tau_J}_{\mathfrak{su}(N)} \propto f_{IK}{}^L f_{JL}{}^K$ is used to raise and lower the indices $I,J,\dots$ which count the generators of $SU(N)$, where $f{}_{IJ}{}^K$ are the structure constants for the $\mathfrak{su}(N)$ algebra. In addition to the topological $B\wedge F$ term, we also have two simplicity constraint terms, featuring the Lagrange multiplier $2$-form $\lambda^I$ and the Lagrange multiplier $0$-form $\zeta^{abI}$. The $0$-form $M_{abI}$ is also a Lagrange multiplier, while $g$ is the coupling constant for the Yang-Mills theory.

Finally, $\delta^a$ is a nondynamical $1$-form, such that there exists a global coordinate frame in which its components are equal to the Kronecker symbol $\delta^a{}_{\mu}$ (hence the notation $\delta^a$). The 1-form $\delta^a$ plays the role of a background field, and defines the global spacetime metric, via the equation
\begin{equation} \label{eq:flatspacetimemetric}
\eta_{\mu\nu} = \eta_{ab} \delta^a{}_{\mu} \delta^b{}_{\nu}\,,
\end{equation}
where $\eta_{ab} \equiv \diag (-1,+1,+1,+1)$ is the Minkowski metric. Since the coordinate system is global, the spacetime manifold $\cM_4$ is understood to be flat. The indices $a,b,\dots$ are local Lorentz indices, taking values $0,\dots,3$. Note that the field $\delta^a$ has all the properties of the tetrad $1$-form $e^a$ in the flat Minkowski spacetime. Also note that the action (\ref{eq:bfgauge}) is manifestly diffeomorphism invariant and gauge invariant with respect to $SU(N)$, but not background independent, due to the presence of $\delta^a$.

The equations of motion are obtained by varying the action (\ref{eq:bfgauge}) with respect to the variables ${\zeta^{ab}}{}^I$, ${M_{ab}}{}_I$, $A^I$,  $B_I$, and $\lambda^I$, respectively (note that we do not take the variation of the action with respect to the background field $\delta^a$):
\begin{gather}
\label{eq:g1}{M_{ab}}_I\varepsilon_{cdef}\delta^c \wedge \delta^d \wedge \delta^e \wedge \delta^f- F_I \wedge \delta_a \wedge \delta_b=0\,, \vphantom{\ds\int} \\
\label{eq:g2}-\frac{12}{g}\lambda^I\wedge \delta^a \wedge \delta^b + \zeta^{ab}{}^I\varepsilon_{cdef}\delta^c \wedge \delta^d \wedge \delta^e \wedge \delta^f=0\,, \vphantom{\ds\int} \\
\label{eq:g3}
-\D B_I+{f}_{JI}{}^K B_K\wedge A^J+\D(\zeta^{ab}{}_I \delta_a \wedge \delta_b)-{f}_{JI}{}^K \zeta^{ab}{}_K \delta_a \wedge \delta_b \wedge A^J=0\,, \vphantom{\ds\int} \\
\label{eq:g4}F_I+\lambda_I=0\,, \vphantom{\ds\int} \\
\label{eq:g5}B_I-\frac{12}{g}{M_{ab}}_I\delta^a\wedge \delta^b=0\,, \vphantom{\ds\int}
\end{gather}
From the algebraic equations (\ref{eq:g1}), (\ref{eq:g2}), (\ref{eq:g4}) and (\ref{eq:g5}) one obtains the multipliers as functions of the dynamical field $A^I$:
\begin{equation}
    M_{ab}{}_I=\frac{1}{48}\varepsilon_{abcd}F{}_I{}^{cd}\,, \quad   {\zeta^{ab}}{}^I=\frac{1}{4g}\varepsilon^{abcd}F{}^I{}_{cd}\,, \quad 
    \lambda{}_I{}_{ab}=F{}_I{}_{ab}\,, \quad
    B{}_I{}_{ab}=\frac{1}{2g}\varepsilon_{abcd}F{}_I{}^{cd}\,.  
\end{equation}
Here we used the notation $F_I{}_{ab}=F_I{}_{\mu\nu}\delta_a{}^\mu \delta_b{}^\nu$, where we used the fact that $\delta^a{}_\mu$ is invertible, and similarly for other variables. Using these equations and the differential equation (\ref{eq:g3}) one obtains the equation of motion for gauge field $A^I$,
\begin{equation}\label{eq:g8}
    \nabla_\rho F^{I\rho \mu}\equiv \partial_\rho F^{I\rho \mu} + f_{JK}{}^I A^J{}_\rho F^K{}^{\rho\mu}=0\,.
\end{equation}
This is precisely the classical equation of motion for the free Yang-Mills theory. Note that in addition to the Yang-Mills theory, one can easily extend the action (\ref{eq:bfgauge}) in order to describe the massive vector field and obtain the Proca equation of motion. This is done by adding a mass term
\begin{equation}
-\frac{1}{4!} m^2 A_{I\mu} A^I{}_{\nu} \eta^{\mu\nu} \lc_{abcd} \delta^a \wedge \delta^b \wedge \delta^c \wedge \delta^d
\end{equation}
to the action (\ref{eq:bfgauge}). Of course, this term explicitly breaks the $SU(N)$ gauge symmetry of the action.

Another example of the constrained $BF$ theory is the Plebanski action for general relativity \cite{plebanski1977}, see also \cite{BFgravity2016} for a recent review. Starting from a gauge group $SO(3,1)$, one constructs a constrained $BF$ action as
\begin{equation}
    S =\int_{\cM_4} B_{ab} \wedge R^{ab} + \phi_{abcd} B^{ab} \wedge B^{cd}\,.
\end{equation}
Here $R^{ab}$ is the curvature $2$-form for the spin connection $\omega^{ab}$, $B_{ab}$ is the usual Lagrange multiplier $2$-form, while $\phi_{abcd}$ is the Lagrange multiplier $0$-form corresponding to the simplicity constraint term $B^{ab}\wedge B^{cd}$. It can be shown that the variation of this action with respect to $B_{ab}$, $\omega^{ab}$ and $\phi_{abcd}$ gives rise to equations of motion which are equivalent to vacuum general relativity. However, the tetrad fields appear in the model as a solution to the simplicity constraint equation of motion $B^{ab}\wedge B^{cd} = 0$. Thus, being intrinsically on-shell objects, they are not present in the action and cannot be quantized. This renders the Plebanski model unsuitable for coupling of matter fields to gravity \cite{RovelliSpinfoamFermions,MikovicVojinovic2012,VojinovicCDT2016}. Nevertheless, as a model for pure gravity, the Plebanski model has been successfully quantized in the context of spinfoam models, see \cite{EPRL,FK,RovelliBook,RovelliVidottoBook} for details and references.

\subsection{$2BF$ theory}\label{subBII}

In order to circumvent the issue of coupling of matter fields, a recent promising approach has been developed \cite{MikovicVojinovic2012,MikovicStrongWeak,MikovicOliveira2014,MOV2016,VojinovicCDT2016,MOV2019} in the context of higher category theory \cite{BaezHuerta2011}. In particular, one employs the higher category theory construction to generalize the $BF$ action to the so-called $2BF$ action, by passing from the notion of a gauge group to the notion of a gauge $2$-group. In order to introduce it, let us first give a short review of the $2$-group formalism.

In the framework of category theory, the group as an algebraic structure can be understood as a specific type of category, namely a category with only one object and invertible morphisms \cite{BaezHuerta2011}. The notion of a category can be generalized to the so-called {\em higher categories}, which have not only objects and morphisms, but also $2$-morphisms (morphisms between morphisms), and so on. This process of generalization is called the {\em categorical ladder}. Similarly to the notion of a group, one can introduce a $2$-group as a $2$-category consisting of only one object, where all the morphisms and $2$-morphisms are invertible. It has been shown that every strict $2$-group is equivalent to a crossed module $(H \stackrel{\del}{\to}G \,, \rhd)$, see Appendix \ref{ApendiksA} for definition. Here $G$ and $H$ are groups, $\delta$ is a homomorphism from $H$ to $G$, while $\rhd:G\times H \to H$ is an action of $G$ on $H$.

An important example of this structure is a vector space $V$ equipped with an isometry group $O$. Namely, $V$ can be regarded as an Abelian Lie group with addition as a group operation, so that a representation of $O$ on $V$ is an action $\rhd$ of $O$ on the group $V$, giving rise to the crossed module $(V \stackrel{\del}{\to}O \,, \rhd)$, where the homomorphism $\del$ is chosen to be trivial, i.e., it maps every element of $V$ into a unit of $O$. We will make use of this example below to introduce the Poincar\'e $2$-group.

Similarly to the case of an ordinary Lie group $G$ which has a naturally associated notion of a connection $\alpha$, giving rise to a $BF$ theory, the $2$-group structure has a naturally associated notion of a $2$-connection $(\alpha\,,\beta)$, described by the usual $\mathfrak{g}$-valued $1$-form $\alpha \in \cA^1(\cM_4\,,\mathfrak{g})$ and an $\mathfrak{h}$-valued $2$-form $\beta \in \cA^2(\cM_4\,,\mathfrak{h})$, where $\mathfrak{h}$ is a Lie algebra of the Lie group $H$. The $2$-connection gives rise to the so-called {\em fake $2$-curvature} $(\cF,\cG)$, given as
\begin{equation}\label{eq:krivine}
    \cF=\D \alpha+ \alpha \wedge \alpha - \partial\beta\,, \quad \quad \cG= d\beta+\alpha\wedge^\rhd \beta\,.
\end{equation}
Here $\alpha \wedge^\rhd \beta$ means that $\alpha$ and $\beta$ are multiplied as forms using $\wedge$, and simultaneously multiplied as algebra elements using $\rhd$, see Appendix \ref{ApendiksA}. The curvature pair $(\cF,\cG)$ is called fake because of the presence of the $\partial\beta$ term in the definition of $\cF$, see \cite{BaezHuerta2011} for details.

Using these variables, one can introduce a new action as a generalization of the $BF$ action, such that it is gauge invariant with respect to both $G$ and $H$ groups. It is called the $2BF$ action and is defined in the following way \cite{GirelliPfeifferPopescu2008,FariaMartinsMikovic2011}:
\begin{equation}\label{eq:bfcg}
S_{2BF} =\int_{\cM_4} \langle B \wedge \cF \rangle_{\mathfrak{ g}} +  \langle C \wedge \cG \rangle_{\mathfrak{h}} \,,
\end{equation}
where the $2$-form $B \in \cA^2(\cM_4\,, \mathfrak{g})$ and the $1$-form $C\in \cA^1(\cM_4\,,\mathfrak{h})$ are Lagrange multipliers. Also, $\langle \_\,,\_\rangle{}_{\mathfrak{g}}$ and $\langle \_\,,\_\rangle{}_{\mathfrak{h}}$ denote the $G$-invariant bilinear symmetric nondegenerate forms for the algebras $\mathfrak{g}$ and $\mathfrak{h}$, respectively. As a consequence of the axiomatic structure of a crossed module (see Appendix \ref{ApendiksA}), the bilinear form $\langle \_\,,\_\rangle{}_{\mathfrak{h}}$ is $H$-invariant as well. See \cite{GirelliPfeifferPopescu2008,FariaMartinsMikovic2011} for review and references.

Similarly to the $BF$ action, the $2BF$ action is also topological, which can be seen from equations of motion. Varying with respect to $B$ and $C$ one obtains
\begin{equation}\label{eq:jedn2bf}
    \cF=0\,, \quad \quad \cG=0\,,
\end{equation}
while varying with respect to $\alpha$ and $\beta$ one obtains the equations for the multipliers,
\begin{gather}
\D B_\alpha- {g_{\alpha \beta}}^\gamma B_\gamma \wedge  \alpha^\beta - {\rhd_{\alpha a}}^b C_b\wedge \beta^a =0\,, \\ 
\D C_a - {\partial_a}^\alpha B_\alpha + {\rhd_{\alpha a}}^b C_b \wedge \alpha^\alpha =0\,.
\end{gather}
One can either show that these equations have only trivial solutions, or one can use the Hamiltonian analysis to show that there are no local propagating degrees of freedom (see for example \cite{MikovicOliveira2014,MOV2016}), demostrating the topological nature of the theory.

An example of a $2$-group relevant for physics is the Poincar\' e $2$-group, which is constructed using the aforementioned example of a vector space equipped with an isometry group. One constructs a crossed module by choosing
\begin{equation}
G=SO(3,1)\,, \qquad H=\realni^4\,,
\end{equation}
while $\rhd$ is a natural action of $SO(3,1)$ on $\realni^4$, and the map $\partial$ is trivial. The $2$-connection $(\alpha, \beta)$ is given by the algebra-valued differential forms
\begin{equation}
\alpha=\omega^{ab}M_{ab}\,, \qquad \beta = \beta^a P_a\,,
\end{equation}
where $\omega^{ab}$ is the spin connection, while $M_{ab}$ and $P_a$ are the generators of groups $SO(3,1)$ and $\realni^4$, respectively. The corresponding $2$-curvature in this case is given by
\begin{equation}
{\cal F} = (\mathrm{d} \omega^{ab} + {\omega^a}_c\wedge\omega^{cb} )M_{ab} \equiv R^{ab}M_{ab} \,,\quad {\cal G} = (\mathrm{d}\beta^a + {\omega^a}_b \wedge \beta^b)P_a \equiv \nabla\beta^a P_a \equiv G^a P_a\,, \label{eq:krivinezapoenc}
\end{equation}
where we have evaluated $\wedge^\rhd$ using the equation $M_{ab}\rhd P_c = \eta_{[bc} P_{a]}$. Note that, since $\del$ is trivial, the fake curvature is the same as ordinary curvature. Using the bilinear forms
\begin{equation}
\killing{M_{ab}}{M_{cd}}_{\mathfrak{g}} = \eta_{a[c} \eta_{bd]} \,, \qquad \killing{P_a}{P_b}_{\mathfrak{h}} = \eta_{ab}\,,
\end{equation}
one can show that $1$-forms $C^a$ transform in the same way as the tetrad $1$-forms $e^a$ under the Lorentz transformations and diffeomorphisms, so the fields $C^a$ can be identified with the tetrads. Then one can rewrite the $2BF$ action (\ref{eq:bfcg}) for the Poincar\'e $2$-group as
\begin{equation}\label{eq:GravityTopoloski}
S_{2BF} = \int_{\cM_4} B^{ab}\wedge R_{ab} + e_a \wedge \nabla\beta^a \,. 
\end{equation}

In order to obtain general relativity, the topological action (\ref{eq:GravityTopoloski}) can be modified by adding a convenient simplicity constraint, like it is done in the $BF$ case:
\begin{equation}\label{eq:GravityVeza}
  S = \int_{\cM_4} B^{ab}\wedge R_{ab} + e_a \wedge \nabla\beta^a - \lambda_{ab} \wedge \Big( B^{ab}-\frac{1}{16\pi l_p^2}\varepsilon^{abcd} e_c \wedge e_d \Big) \,.
\end{equation}
Here $\lambda_{ab}$ is a Lagrange multiplier $2$-form associated to the simplicity constraint term, and $l_p$ is the Planck length. Varying the action (\ref{eq:GravityVeza}) with respect to $B_{ab}$, $e_{a}$, $\omega_{ab}$, $\beta_{a}$ and $\lambda_{ab}$, one obtains the following equations of motion:
\begin{gather}  
\label{eq:g3a}
R_{ab} - \lambda_{ab} = 0\,,\vphantom{\ds\int}\\
\label{eq:g4a}
\nabla \beta_a + \frac{1}{8\pi l_p^2} \varepsilon_{abcd} \lambda^{bc} \wedge e^d=0\,,\vphantom{\ds\int}\\
\label{eq:g5a}
\nabla B_{ab} - e_{[a} \wedge \beta_{b]} = 0\,,\vphantom{\ds\int}\\
\label{eq:g6a}
\nabla e_a = 0\,,\vphantom{\ds\int}\\
\label{eq:g7a}
B^{ab}-\frac{1}{16\pi l_p^2}\varepsilon^{abcd} e_c \wedge e_d=0\,.\vphantom{\ds\int}
\end{gather}
The only dynamical fields are the tetrads $e^a$, while all other fields can be algebraically determined, as follows. From the equations (\ref{eq:g6a}) and (\ref{eq:g7a}) we obtain that $\nabla B^{ab} = 0$, from which it follows, using the equation (\ref{eq:g5a}), that $e_{[a} \wedge \beta_{b]} = 0$. Assuming that the tetrads are nondegenerate, $e\equiv \det (e^a{}_{\mu}) \neq 0$, it can be shown that this is equivalent to the condition $\beta^{a}=0$ (for the proof see Appendix in \cite{MikovicVojinovic2012}). Therefore, from the equations (\ref{eq:g3a}), (\ref{eq:g5a}), (\ref{eq:g6a}) and (\ref{eq:g7a}) we obtain
\begin{equation}
    \lambda^{ab}{}_{\mu\nu}=R^{ab}{}_{\mu\nu}\,,  \quad
    \beta^a{}_{\mu\nu}=0\,, \quad
    B_{ab}{}_{\mu \nu}=\frac{1}{8\pi l_p^2}\varepsilon_{abcd} e^c{}_\mu e^d{}_\nu\,, \quad
    \omega^{ab}{}_\mu =\triangle^{ab}{}_\mu\,.
\end{equation}
Here the Ricci rotation coefficients are defined as
\begin{equation}
    \triangle^{ab}{}_\mu\equiv \frac{1}{2}(c^{abc}-c^{cab}+c^{bca})e_{c\mu}\,, 
\end{equation}
where
\begin{equation}
    c^{abc}= e^{\mu}{}_b e^{\nu}{}_c \left( \partial_\mu e^a{}_\nu - \partial_\nu e^a{}_\mu \right)\,.
\end{equation}
Finally, the remaining equation (\ref{eq:g4a}) reduces to
\begin{equation}
 \varepsilon_{abcd} R^{bc} \wedge e^d=0\,,
\end{equation}
which is nothing but the vacuum Einstein field equation $R_{\mu\nu}-\frac{1}{2}g_{\mu\nu}R = 0$. Therefore, the action (\ref{eq:GravityVeza}) is classically equivalent to general relativity.

The main advantage of the action (\ref{eq:GravityVeza}) over the Plebanski model and similar approaches lies in the fact that the tetrad fields are explicitly present in the topological sector of the theory. This allows one to couple matter fields in a straightforward way, as demonstrated in \cite{MikovicVojinovic2012}. However, one can do even better, and couple gauge fields to gravity within a unified framework of $2$-group formalism.

Let us demonstrate this on the example of the $SU(N)$ Yang-Mills theory. Begin by modifying the Poincar\'e $2$-group structure to include the $SU(N)$ gauge group, as follows. We choose the two Lie groups as
\begin{equation}
G=SO(3,1)\times SU(N)\,, \qquad H=\mathbb{R}^4\,,
\end{equation}
and we define the action $\rhd$ of the group $G$ in the following way. As in the case of the Poincar\'e $2$-group, it acts on itself via conjugation. Next, it acts on $H$ such that the $SO(3,1)$ subgroup acts on $\realni^4$ via the vector representation, while the action of $SU(N)$ subgroup is trivial. The map $\partial$ also remains trivial, as before. The $2$-connection $(\alpha,\beta)$ now obtains the form which reflects the structure of the group $G$,
\begin{equation}
\alpha = \omega^{ab}M_{ab} + A^I \tau_I\,, \qquad \beta = \beta^a P_a\,,
\end{equation}
where $A^I$ is the gauge connection $1$-form, while $\tau_I$ are the $SU(N)$ generators. The curvature for $\alpha$ is thus
\begin{equation}
{\cal F} = R^{ab}M_{ab} + F^I \tau_I\,, \qquad F^I \equiv \rmd A^I + f_{JK}{}^I A^J \wedge A^K\,.
\end{equation}
The curvature for $\beta$ remains the same as before, since the action $\rhd$ of $SU(N)$ on $\realni^4$ is trivial, i.e., $\tau_I \rhd P_a = 0$. Finally, the product structure of the group $G$ implies that its Killing form $\killing{\_}{\_}_{\mathfrak{g}}$ reduces to the Killing forms for the $SO(3,1)$ and $SU(N)$, along with the identity $\killing{M_{ab}}{\tau_I}_{\mathfrak{g}}=0$.

Given a crossed module defined in this way, its corresponding topological $2BF$ action (\ref{eq:bfcg}) becomes
\begin{equation} \label{eq:bfcggauge}
    S_{2BF}=\int_{\cM_4} B^{ab}\wedge R_{ab} + B^I \wedge F_I + e_a \wedge \nabla \beta^a\,,
\end{equation}
where $B^I \in \cA^2(\cM_4\,,\mathfrak{su}(N))$ is the new Lagrange multiplier. In order to transform this topological action into action with nontrivial dynamics, we again introduce the appropriate simplicity constraints. The constraint giving rise to gravity is the same as in (\ref{eq:GravityVeza}), while the constraint for the gauge fields is given as in the action (\ref{eq:bfgauge}) with the substitution $\delta^a \to e^a$:
\begin{equation} \label{eq:YMplusGravity}
\begin{aligned}
    S=&\int_{\cM_4} B^{ab}\wedge R_{ab} + B^I \wedge F_I + e_a \wedge \nabla \beta^a - \lambda_{ab} \wedge \Big(B^{ab}-\frac{1}{16\pi l_p^2}\varepsilon^{abcd} e_c \wedge e_d\Big) \\ &+ \lambda^I\wedge \Big(B_I-\frac{12}{g}{M_{ab}}_Ie^a\wedge e^b\Big) + {\zeta^{ab}}{}^I \Big( {M_{ab}}{}_I\varepsilon_{cdef}e^c\wedge e^d \wedge e^e \wedge e^f- g_{IJ}F^J \wedge e_a \wedge e_b \Big) \,.
\end{aligned}
\end{equation}
It is crucial to note that the action (\ref{eq:YMplusGravity}) is a combination of the pure gravity action (\ref{eq:GravityVeza}) and the Yang-Mills action (\ref{eq:bfgauge}), such that the nondynamical background field $\delta^a$ from (\ref{eq:bfgauge}) gets promoted to a dynamical field $e^a$. The relationship between these fields has already been hinted at in the equation (\ref{eq:flatspacetimemetric}), which describes the connection between $\delta^a$ and the flat spacetime metric $\eta_{\mu\nu}$. Once promoted to $e^a$, this field becomes dynamical, while the equation (\ref{eq:flatspacetimemetric}) becomes the usual relation between the tetrad and the metric,
\begin{equation}
g_{\mu\nu} = \eta_{ab} e^a{}_{\mu} e^b{}_{\nu}\,,
\end{equation}
further confirming that the Lagrange multiplier $C^a$ should be identified with the tetrad. Moreover, the total action (\ref{eq:YMplusGravity}) now becomes background independent, as expected in general relativity. All this is a consequence of the fact that the tetrad field is explicitly present in the topological sector of the action (\ref{eq:GravityVeza}), establishing an improvement over the Plebanski model.

By varying the action (\ref{eq:YMplusGravity}) with respect to the variables $B_{ab}$, $\omega_{ab}$, $\beta_a$, $\lambda_{ab}$, ${\zeta^{ab}}{}^I$, ${M_{ab}}{}_I$, $B_I$, $\lambda^I$, $A^I$, and $e^a$, we obtain the following equations of motion, respectively:
\begin{gather}
\label{eq:01}R^{ab}-\lambda^{ab}=0\,, \vphantom{\ds\int} \\
\label{eq:02}\nabla B^{ab} - e^{[a} \wedge \beta^{b]} = 0\,, \vphantom{\ds\int} \\
\label{eq:03}\nabla e^a = 0\,,  \vphantom{\ds\int} \\
\label{eq:04}B_{ab}-\frac{1}{16\pi l_p^2}\varepsilon_{abcd} e^c \wedge e^d=0\,, \vphantom{\ds\int} \\
\label{eq:05}{M_{ab}}_I\varepsilon_{cdef}e^c \wedge e^d \wedge e^e \wedge e^f- F_I \wedge e_a \wedge e_b=0\,, \vphantom{\ds\int} \\
\label{eq:06}-\frac{12}{g}\lambda^I\wedge e^a \wedge e^b + \zeta^{abI} \varepsilon_{cdef}e^c \wedge e^d \wedge e^e \wedge e^f=0\,, \vphantom{\ds\int} \\
\label{eq:09}F_I+\lambda_I=0\,, \vphantom{\ds\int} \\
\label{eq:010}B_I-\frac{12}{g}{M_{ab}}_Ie^a\wedge e^b=0\,, \vphantom{\ds\int} \\
\label{eq:08}
-\D B_I+B_K\wedge g_{JI}{}^K A^J+\D(\zeta^{ab}_I e_a \wedge e_b)- \zeta^{ab}_K e_a \wedge e_b \wedge g_{JI}{}^K A^J=0\,, \vphantom{\ds\int} \\
\nabla \beta_a + \frac{1}{8\pi l_p^2} \varepsilon_{abcd} \lambda^{bc} \wedge e^d- \frac{24}{g}{M_{ab}}_I\lambda^I\wedge e^b \nonumber \\
\label{eq:011} + 4{\zeta^{ef}}^I{M_{ef}}_I\varepsilon_{abcd}e^b\wedge e^c \wedge e^d - 2 {\zeta_{ab}}^I F_I\wedge e^b=0\,. \vphantom{\ds\int}
\end{gather}
In the above system of equations, we have two dynamical equations for $e^a$ and $A^I$, while all other variables are algebraically determined from these. In particular, from equations (\ref{eq:01})--(\ref{eq:010}), we have:
\begin{equation}\label{eq:012}
    \begin{aligned}
        \lambda_{ab}{}_{\mu\nu}=R_{ab}{}_{\mu\nu}\,,\quad  \beta_a{}_{\mu \nu}=0\,,\quad \omega_{ab}{}_\mu&=\triangle_{ab}{}_\mu\,, \quad 
    \lambda{}_{ab}{}_I=F_{ab}{}_I\,, \quad
    B_{\mu\nu}{}_I=-\frac{e}{2g}{\varepsilon_{\mu\nu\rho\sigma}} {F^{\rho \sigma}}_I\,,\vphantom{\ds\int}\\
    B_{ab}{}_{\mu \nu}=\frac{1}{8\pi l_p^2}\varepsilon_{abcd} e^c{}_\mu e^d{}_\nu\,,\quad  M_{ab}{}_I=&-\frac{1}{4eg}\varepsilon^{\mu \nu \rho \sigma}F_{\mu \nu}{}^Ie^a{}_\rho e^b{}_\sigma\,, \quad  {\zeta^{ab}}{}^I=\frac{1}{4eg}\varepsilon^{\mu \nu \rho \sigma}F_{\mu \nu}{}^Ie^a{}_\rho e^b{}_\sigma\,.
    \end{aligned}
\end{equation}
Then, substituting all these into (\ref{eq:08}) and (\ref{eq:011}) we obtain the differential equation of motion for $A^I$,
\begin{equation}
   \nabla_\rho F^{I \rho \mu}\equiv  \partial_\rho F^{I\rho\mu} + \itGamma^{\rho}{}_{\lambda\rho} F^{I\lambda\mu} + f_{JK}{}^I A^J{}_\rho F^{K\rho\mu} =0\,,\label{eq:020}
\end{equation}
where $\itGamma^{\lambda}{}_{\mu\nu}$ is the standard Levi-Civita connection, and a differential equation of motion for $e^a$,
\begin{equation}
   R^{\mu \nu}-\frac{1}{2}g^{\mu \nu}R=8\pi l_p^2 \; T^{\mu \nu}\,, \qquad
T^{\mu \nu} \equiv -\frac{1}{4g}\left(F_{\rho \sigma}{}^IF^{\rho \sigma}{}_Ig^{\mu \nu}+4F^{\mu \rho}{}_I{F_\rho}^{\nu}{}^I \right)\,.\label{eq:019}
\end{equation}
The system of equations (\ref{eq:012})--(\ref{eq:019}) is equivalent to the system (\ref{eq:01})--(\ref{eq:011}). Note that we have again obtained that $\beta^a=0$, as in the pure gravity case.

In this way, we see that both gravity and gauge fields can be represented within a unified framework of higher gauge theory based on a $2$-group structure.

\section{\label{SecIII}$3BF$ models, scalar and fermion matter fields}

While the structure of a $2$-group can successfully accommodate both gravitational and gauge fields, unfortunately it cannot include other matter fields, such as scalars or fermions. In order to construct a unified description of all matter fields within the framework of higher gauge theory, we are led to make a further generalization, passing from the notion of a $2$-group to the notion of a $3$-group. As it turns out, the $3$-group structure is a perfect fit for the description of all fields that are present in the Standard Model, coupled to gravity. Moreover, this structure gives rise to a new gauge group, which corresponds to the choice of the scalar and fermion fields present in the theory. This is a novel and unexpected result, which has the potential to open up a new avenue of research with the aim of explaining the structure of the matter sector of the Standard Model and beyond.

In order to demonstrate this in more detail, we first need to introduce the notion of a $3$-group, which we will afterward use to construct constrained $3BF$ actions describing scalar and fermion fields on an equal footing with gravity and gauge fields.

\subsection{$3$-groups and topological $3BF$ action}\label{sec:III.1}

Similarly to the concepts of a group and a $2$-group, one can introduce the notion of a $3$-group in the framework of higher category theory, as a $3$-category with only one object where all the morphisms, $2$-morphisms and $3$-morphisms are invertible. It has been proved that a strict $3$-group is equivalent to a {\em $2$-crossed module} \cite{martins2011}, in the same way as a $2$-group is equivalent to a crossed module.

A Lie $2$-crossed module, denoted as $(L\stackrel{\delta}{\to} H \stackrel{\partial}{\to}G\,, \rhd\,, \{\_\,,\_\})$, is a algebraic structure specified by three Lie groups $G$, $H$ and $L$, together with the homomorphisms $\delta$ and $\del$, an action $\rhd$ of the group $G$ on all three groups, and a $G$-equivariant map
\begin{displaymath}
\{ \_ \,, \_ \} : H\times H \to L\,.
\end{displaymath}
called the Peiffer lifting. See Appendix \ref{ApendiksA} for more details.

In complete analogy to the construction of $BF$ and $2BF$ topological actions, one can define a gauge invariant topological $3BF$ action for the manifold $\mathcal{M}_4$ and $2$-crossed module $(L\stackrel{\delta}{\to} H \stackrel{\partial}{\to}G\,, \rhd\,, \{\_\,,\_\})$. Given $\mathfrak{g}$, $\mathfrak{h}$ and $\mathfrak{l}$ as Lie algebras corresponding to the groups $G$, $H$ and $L$, one can introduce a $3$-connection $(\alpha, \beta,\gamma)$ given by the algebra-valued differential forms $\alpha \in \cA^1(\cM_4\,,\mathfrak{g})$, $\beta \in \cA^2(\cM_4\,,\mathfrak{h})$ and $\gamma \in \cA^3(\cM_4\,,\mathfrak{l})$. The corresponding fake $3$-curvature $(\cal F\,, G\,, H)$ is then defined as
\begin{equation}\label{eq:3krivine}
    \cF = \D \alpha+\alpha \wedge \alpha - \partial \beta \,, \quad \quad \cG = \D \beta + \alpha \wedge^\rhd \beta - \delta \gamma\,, \quad \quad \cH = \D \gamma + \alpha\wedge^\rhd \gamma + \{\beta \wedge \beta\} \,.
    \end{equation}
see \cite{martins2011, Wang2014} for details. Then, a $3BF$ action is defined as
\begin{equation}\label{eq:bfcgdh}
S_{3BF} =\int_{\mathcal{M}_4} \langle B \wedge  {\cal F} \rangle_{\mathfrak{ g}} +  \langle C \wedge  {\cal G} \rangle_{\mathfrak{h}} + \langle D \wedge {\cal H} \rangle_{\mathfrak{l}} \,,
\end{equation}
where $B \in \cA^2(\cM_4,\mathfrak{g})$, $C \in \cA^1(\cM_4,\mathfrak{h})$ and $D \in \cA^0(\cM_4,\mathfrak{l})$ are Lagrange multipliers. The forms $\killing{\_}{\_}_{\mathfrak{g}}$, $\killing{\_}{\_}_{\mathfrak{h}}$ and $\killing{\_}{\_}_{\mathfrak{l}}$ are $G$-invariant bilinear symmetric nondegenerate forms on $\mathfrak{g}$,  $\mathfrak{h}$ and $\mathfrak{l}$, respectively. Under certain conditions, the forms $\killing{\_}{\_}_{\mathfrak{h}}$ and $\killing{\_}{\_}_{\mathfrak{l}}$ are also $H$-invariant and $L$-invariant, see Appendix \ref{ApendiksB} for details.

One can see that varying the action with respect to the variables $B$, $C$ and $D$, one obtains the equations of motion 
\begin{equation}\label{eq:jed3bf}
\mathcal{F} = 0 \,, \quad \quad \mathcal{G} = 0\,, \quad \quad \mathcal{H} = 0 \,,
\end{equation}
while varying with respect to $\alpha$, $\beta$, $\gamma$ one obtains
\begin{gather}
\D B_\alpha- {g_{\alpha \beta}}^\gamma B_\gamma \wedge  \alpha^\beta - {\rhd_{\alpha a}}^b C_b\wedge \beta^a+\rhd_{\alpha B}{}^A D_A\wedge\gamma^B=0\,, \\ 
\D C_a - {\partial_a}^\alpha B_\alpha + {\rhd_{\alpha a}}^b C_b \wedge \alpha^\alpha + 2X_{\{ab\}}{}^AD_A\wedge \beta^b=0\,,\\
\D D_A - \rhd_{ \alpha A}{}^B D_B\wedge \alpha^\alpha+\delta_A {}^a C_a=0\,.
\end{gather}

Regarding the gauge transformations, the $3BF$ action is invariant with respect to three different types of transformations, generated by the groups $G$, $H$ and $L$, respectively. Under the $G$-gauge transformations, the $3$-connection transforms as
\begin{equation}\label{eq:conGg}
    \alpha'= g^{-1}\alpha g + g^{-1}\D g\,, \quad \quad \beta'= g^{-1}\rhd \beta\,, \quad\quad \quad \gamma' = g^{-1} \rhd \gamma\,,
\end{equation}
where $g:\cM_4 \to G$ is an element of the $G$-principal bundle over $\cM_4$. Next, under the $H$-gauge transformations, generated by $\eta \in \cA^1 (\cM_4\,,\mathfrak{h})$, the $3$-connection transforms as
\begin{equation}\label{eq:conHg}
    \alpha' = \alpha + \partial\eta\, , \quad \quad \beta' = \beta + \D \eta + \alpha' \wedge^{ \rhd} \eta - \eta \wedge \eta\, ,  \quad \quad \gamma' = \gamma - \{\beta'\, \wedge \, \eta\} - \{ \eta \, \wedge \, \beta\}\,. 
\end{equation}
Finally, under the $L$-gauge transformations, generated by $\theta\in\cA^2(\cM_4\,,\mathfrak{l})$, the $3$-connection transforms as
\begin{equation}\label{eq:conLg}
    \alpha' = \alpha\,, \quad \quad \quad \beta' = \beta - \delta\theta\,, \quad \quad \quad \gamma' = \gamma - \D \theta - \alpha \wedge \theta\,.
\end{equation}
As a consequence of the definition (\ref{eq:3krivine}) and the above transformation rules, the curvatures transform under the $G$-gauge transformations as
\begin{equation}
    \cF \to g^{-1} \cF g\,, \quad \quad \quad \cG \to g^{-1}\rhd \cG\,, \quad  \quad \quad \cH \to g^{-1}\rhd \cH\,,
\end{equation}
under the $H$-gauge transformations as
\begin{equation}
    \cF \to \cF\, , \quad \quad \quad \cG \to \cG + \cF \wedge^\rhd \eta\,, \quad \quad \quad \cH \to \cH - \, \{ \cG'\, \wedge \, \eta\} + \{ \eta \, \wedge \, \cG \}\,, 
\end{equation}
and under the $L$-gauge transformations as
\begin{equation}
    \cF \to \cF\, , \quad \quad \cG \to \cG\,,  \quad \quad \quad \cH \to \cH - \cF \wedge^\rhd \theta\,.
\end{equation}
For more details, the reader is referred to \cite{Wang2014}.

In order to make the action (\ref{eq:bfcgdh}) gauge invariant with respect to the transformations (\ref{eq:conGg}), (\ref{eq:conHg}) and (\ref{eq:conLg}), the Lagrange multipliers $B$, $C$ and $D$ must transform under the $G$-gauge transformations as
\begin{equation}\label{eq:LGg}
    B \to g^{-1} B g\,, \quad \quad \quad C \to g^{-1}\rhd C\,, \quad  \quad \quad D \to g^{-1}\rhd D\,,
\end{equation}
under the $H$-gauge transformations as
\begin{equation}\label{eq:LHg}
    B \to B + C'\wedge^{\cT}\eta-\eta\wedge^{\cal D}\eta\wedge^{\cal D}D\,, \quad \quad \quad C \to C+D\wedge^{{\cal X}_1}\eta+D\wedge^{{\cal X}_2}\eta\,, \quad \quad \quad D \to D\,,
\end{equation}
while under the $L$-gauge transformations they transform as
\begin{equation}\label{eq:LLg}
    B \to B-D \wedge^{\cal S}\theta\,, \quad \quad \quad C \to C\,, \quad \quad \quad D \to D\,.
\end{equation}
See Appendix \ref{ApendiksB} for details, for the definition of the maps $\cT$, $\cD$, $\cX_1$, $\cX_2$, $\cS$, and for the notation of the $\wedge^{\cT}$, $\wedge^{\cD}$, $\wedge^{\cX_1}$, $\wedge^{\cX_2}$, and $\wedge^{\cS}$ products.

\bigskip

\subsection{Constrained $3BF$ action for a real Klein-Gordon field}\label{sec:III.2}

Once the topological $3BF$ action is specified, we can proceed with the construction of the constrained $3BF$ action, describing a realistic case of a scalar field coupled to gravity. In order to perform this construction, we have to define a specific $2$-crossed module which gives rise to the topological sector of the action, and then we have to impose convenient simplicity constraints. 

We begin by defining a $2$-crossed module $(L\stackrel{\delta}{\to} H \stackrel{\partial}{\to}G\,, \rhd\,, \{\_\,,\_\})$, as follows. The groups are given as
\begin{equation}
G=SO(3,1)\,, \quad \quad H=\mathbb{R}^4\,, \quad \quad L=\mathbb{R}\,.
\end{equation}
The group $G$ acts on itself via conjugation, on $H$ via the vector representation, and on $L$ via the trivial representation. This specifies the definition of the action $\rhd$. The map $\partial$ is chosen to be trivial, as before. The map $\delta$ is also trivial, that is, every element of $L$ is mapped to the identity element of $H$. Finally, the Peiffer lifting is trivial as well, mapping every ordered pair of elements in $H$ to an identity element in $L$. This specifies one concrete $2$-crossed module.

Given this choice of a $2$-crossed module, the $3$-connection $(\alpha\,,\beta\,,\gamma)$ takes the form
\begin{equation}
 \alpha = \omega^{ab}M_{ab}\,, \quad \quad \beta=\beta^a P_a\,, \quad \quad \gamma = \gamma \mathbb{I}\,,
\end{equation}
where $\mathbb{I}$ is the sole generator of the Lie group $\realni$. From (\ref{eq:3krivine}), the fake $3$-curvature $(\cF\,, \cG\,, \cH)$ reduces to the ordinary $3$-curvature,
\begin{equation}
\cF = R^{ab} M_{ab}\,, \quad \quad \cG= \nabla \beta^a P_a\,, \quad \quad \cH= \D \gamma\,,
\end{equation}
where we used the fact that $G$ acts trivially on $L$, that is, $M_{ab}\rhd \mathbb{I}=0$. The topological $3BF$ action (\ref{eq:bfcgdh}) now becomes
\begin{equation}\label{eq:Scalartopoloski}
    S_{3BF}=\int_{\cM_4} B^{ab}\wedge R_{ab} + e_a\wedge \nabla \beta^a + \phi \, \D \gamma\,,
\end{equation}
where the bilinear form for $L$ is $\killing{\mathbb{I}}{\mathbb{I}}_{\mathfrak{l}} = 1$.

It is important to note that the Lagrange multiplier $D$ in (\ref{eq:bfcgdh}) is a $0$-form and transforms trivially with respect to $G$, $H$ and $L$ gauge transformations for our choice of the $2$-crossed module, as can be seen from (\ref{eq:LGg}), (\ref{eq:LHg}) and (\ref{eq:LLg}). Thus, $D$ has all the {\em hallmark properties of a real scalar field}, allowing us to make identification between them, and conveniently relabel $D$ into $\phi$ in (\ref{eq:Scalartopoloski}). This is a crucial property of the $3$-group structure in a $4$-dimensional spacetime and is one of the main results of the paper. It follows the line of reasoning used in recognizing the Lagrange multiplier $C^a$ in the $2BF$ action for the Poincar\'e $2$-group as a tetrad field $e^a$. It is also important to stress that the choice of the third gauge group, $L$, dictates the number and the structure of the matter fields present in the action. In this case, $L=\realni$ implies that we have only one real scalar field, corresponding to a single generator $\mathbb{I}$ of $\realni$. The trivial nature of the action $\rhd$ of $SO(3,1)$ on $\realni$ also implies that $\phi$ transforms as a scalar field. Finally, the scalar field appears as a degree of freedom in the topological sector of the action, making the quantization procedure feasible.

As in the case of $BF$ and $2BF$ theories, in order to obtain nontrivial dynamics, we need to impose convenient simplicity constraints on the variables in the action (\ref{eq:Scalartopoloski}). Since we are interested in obtaining the scalar field $\phi$ of mass $m$ coupled to gravity in the standard way, we choose the action in the form:
\begin{equation}\label{eq:scalar}
\begin{aligned}
 S =\int_{\cM_4} & B^{ab}\wedge R_{ab} + e_a\wedge \nabla \beta^a + \phi \, \D \gamma \vphantom{\ds\int} \\
 &- \lambda_{ab} \wedge \Big(B^{ab}-\frac{1}{16\pi l_p^2}\varepsilon^{abcd} e_c \wedge e_d\Big)\vphantom{\ds\int} \\
 &+ {\lambda}\wedge \Big(\gamma - \frac{1}{2} H_{abc} e^a \wedge e^b \wedge e^c\Big) +\Lambda^{ab}\wedge \Big( H_{abc}\varepsilon^{cdef}e_d\wedge e_e \wedge e_f- \D \phi \wedge e_a \wedge e_b\Big) \vphantom{\ds\int} \\
 &-\frac{1}{2\cdot 4!} m^2\phi^2 \varepsilon_{abcd}e^a\wedge e^b \wedge e^c \wedge e^d\,.\vphantom{\ds\int}
\end{aligned}
\end{equation}
Note that the first row is the topological sector (\ref{eq:Scalartopoloski}), the second row is the familiar simplicity constraint for gravity from the action (\ref{eq:GravityVeza}), the third row contains the new simplicity constraints corresponding to the Lagrange multiplier $1$-forms $\lambda$ and $\Lambda^{ab}$ and featuring the Lagrange multiplier $0$-form $H_{abc}$, while the fourth row is the mass term for the scalar field.

Varying the total action (\ref{eq:scalar}) with respect to the variables $B_{ab}$, $\omega_{ab}$, $\beta_a$, $\lambda_{ab}$, $\Lambda_{ab}$, $\gamma$, ${\lambda}$, $H_{abc}$, $\phi$ and $e^a$ one obtains the equations of motion:
\begin{gather}
\label{eq:31}
R^{ab}-\lambda^{ab}=0\,,\vphantom{\ds\int}\\
\label{eq:32}
\nabla B^{ab} - e^{[a} \wedge \beta^{b]} = 0\,,\vphantom{\ds\int}\\
\label{eq:33}
\nabla e^a = 0\,,\vphantom{\ds\int}\\
\label{eq:34} 
B_{ab}-\frac{1}{16\pi l_p^2}\varepsilon_{abcd} e^c \wedge e^d=0\,,\vphantom{\ds\int}\\
\label{eq:35} 
H_{abc}\varepsilon^{cdef}e_d\wedge e_e \wedge e_f- \D \phi \wedge e_a \wedge e_b=0\,,\vphantom{\ds\int}\\
\label{eq:36} 
\D \phi-{\lambda}=0\,,\vphantom{\ds\int}\\
\label{eq:37} 
\gamma - \frac{1}{2} H_{abc} e^a \wedge e^b \wedge e^c=0\,,\vphantom{\ds\int}\\
\label{eq:38} 
-\frac{1}{2}\lambda\wedge e^a \wedge e^b \wedge e^c+\varepsilon^{cdef}\Lambda^{ab}\wedge e_d\wedge e_e \wedge e_f=0\,,\vphantom{\ds\int}\\
\label{eq:39} 
\D \gamma-\D (\Lambda^{ab} \wedge e_a \wedge e_b)-\frac{1}{4!}  m^2 \phi \varepsilon_{abcd}e^a\wedge e^b \wedge e^c \wedge e^d=0\,,\vphantom{\ds\int}\\
\label{eq:40}
\begin{aligned}
    \nabla \beta_a&+\frac{1}{8\pi l_p^2}\varepsilon_{abcd}\lambda^{bc}\wedge e^d+\frac{3}{2} H_{abc}\lambda\wedge e^b\wedge e^c+3H^{def}\varepsilon_{abcd}\Lambda_{ef}\wedge e^b\wedge e^c\vphantom{\ds\int}\\&-2\Lambda_{ab}\wedge \D \phi \wedge e^b-2\frac{1}{4!}  m^2 \phi \varepsilon_{abcd} e^b \wedge e^c \wedge e^d=0\,.\vphantom{\ds\int}
\end{aligned}
\end{gather}
The dynamical degrees of freedom are $e^a$ and $\phi$, while the remaining variables are algebraically determined in terms of them. Specifically, the equations (\ref{eq:31})--(\ref{eq:38}) give
\begin{equation}
\begin{array}{c}\label{eq:sys1}
\ds    \lambda_{ab}{}_{\mu\nu}=R_{ab}{}_{\mu\nu}\,, \qquad    \omega^{ab}{}_\mu=\triangle^{ab}{}_\mu\,, \qquad 
    \gamma_{\mu\nu\rho}=-\frac{e}{2}\varepsilon_{\mu\nu\rho\sigma}\partial^\sigma\phi\,, \vphantom{\ds\int} \\
\ds    \Lambda^{ab}{}_{\mu}=\frac{1}{12e}g_{\mu\lambda}\varepsilon^{\lambda\nu\rho\sigma}\partial_\nu\phi e{}^a{}_{\rho} e{}^b{}_\sigma\,,\vphantom{\ds\int} \qquad \beta^a{}_{\mu\nu}=0\,, \qquad B_{ab}{}_{\mu\nu}=\frac{1}{8\pi l_p^2}\varepsilon_{abcd}e^c{}_\mu e^d{}_\nu\,, \\
\ds H^{abc}=\frac{1}{6e} \varepsilon^{\mu\nu\rho\sigma}\partial_\mu\phi e^a{}_\nu e^b{}_\rho e^c{}_\sigma\,, \qquad \lambda_{\mu}=\partial_{\mu}\phi\,.\vphantom{\ds\int}
\end{array}
\end{equation}
Note that from the equations (\ref{eq:32}), (\ref{eq:33}) and (\ref{eq:34}) it follows that $\beta^a=0$, as in the pure gravity case. The equation of motion (\ref{eq:39}) reduces to the covariant Klein-Gordon equation for the scalar field,
\begin{equation}
\left(\nabla_\mu\nabla^\mu -m^2\right)\phi=0\,.
\end{equation}
Finally, the equation of motion (\ref{eq:40}) for $e^a$ becomes:
\begin{equation}\label{eq:scalareomfore}
   {R}^{\mu\nu}-\frac{1}{2}g^{\mu\nu} R=8\pi l_p^2 \; T^{\mu\nu}\,, \qquad T^{\mu\nu}\equiv\partial^\mu \phi \partial^\nu \phi -\frac{1}{2}g^{\mu\nu} \left(\partial_\rho \phi \partial^\rho \phi+m^2\phi^2 \right)\,.
\end{equation}
The system of equations (\ref{eq:31})--(\ref{eq:40}) is equivalent to the system of equations (\ref{eq:sys1})--(\ref{eq:scalareomfore}). Note that in addition to the correct covariant form of the Klein-Gordon equation, we have also obtained the correct form of the stress-energy tensor for the scalar field.

\subsection{Constrained $3BF$ action for the Dirac field}\label{sec:III.3}

Now we pass to the more complicated case of the Dirac field. We first define a $2$-crossed module $(L\stackrel{\delta}{\to} H \stackrel{\partial}{\to}G\,, \rhd\,, \{\_\,,\_\})$ as follows. The groups are:
\begin{equation}
G=SO(3,1)\,, \quad \quad H=\mathbb{R}^4\,, \quad \quad L=\mathbb{R}^8(\grasmanovi) \,,
\end{equation}
where $\grasmanovi$ is the algebra of complex Grassmann numbers. The maps $\partial$,  $\delta$ and the Peiffer lifting are trivial. The action of the group $G$ on itself is given via conjugation, on $H$ via vector representation, and on $L$ via spinor representation, as follows. Denoting the $8$ generators of the Lie group $\mathbb{R}^8(\grasmanovi)$ as $P_{\alpha}$ and $P^{\alpha}$, where the index $\alpha$ takes the values $1,\dots,4$, the action of $G$ on $L$ is thus given explicitly as
\begin{equation} \label{eq:actionOfGonLdirac}
M_{ab}\rhd P_{\alpha}=\frac{1}{2}(\sigma_{ab}){}^{\beta}{}_{\alpha} P_{\beta}\,, \qquad M_{ab} \rhd P^{\alpha}=-\frac{1}{2}(\sigma_{ab}){}^{\alpha}{}_{\beta} P^{\beta}\,,
\end{equation}
where $\sigma_{ab}=\frac{1}{4}[\gamma_a,\gamma_b]$, and $\gamma_a$ are the usual Dirac matrices, satisfying the anticommutation rule $\{ \gamma_a\,,\, \gamma_b  \} = -2\eta_{ab}$.

As in the case of the scalar field, the choice of the group $L$ dictates the matter content of the theory, while the action $\rhd$ of $G$ on $L$ specifies its transformation properties. To see this explicitly, let us construct the corresponding $3BF$ action. The $3$-connection $(\alpha\,,\beta\,,\gamma)$ now takes the form
\begin{equation}
\alpha = \omega^{ab}M_{ab}\,, \quad \quad \beta=\beta^a P_a\,, \quad \quad \gamma = \gamma^{\alpha} P_{\alpha}+\bar{\gamma}{}_{\alpha} P^{\alpha}\,,
\end{equation}
while the $3$-curvature $(\cF\,, \cG\,, \cH)$, defined in (\ref{eq:3krivine}), is given as
\begin{equation}
\begin{aligned}
    \cF = R^{ab} M_{ab}\,, \quad & \quad \cG= \nabla \beta^a P_a\,,\\
    \cH= \Big(\D \gamma^{\alpha}+\frac{1}{2}\omega^{ab}(\sigma_{ab}){}^{\alpha}{}_{\beta}\gamma^{\beta}\Big)P_{\alpha}+ \Big(\D \bar{\gamma}{}_{\alpha}-&\frac{1}{2}\omega^{ab}\bar{\gamma}_{\beta}(\sigma_{ab}){}^{\beta}{}_{\alpha}\Big)P^{\alpha}\equiv (\overset{\rightarrow}{\nabla} \gamma){}^{\alpha} P_{\alpha}+ (\bar{\gamma}\overset{\leftarrow}{\nabla}){}_{\alpha}P^{\alpha}\,,
\end{aligned} 
\end{equation}
where we have used (\ref{eq:actionOfGonLdirac}).
The bilinear form $\killing{\_}{\_}_{\mathfrak{l}}$ is defined as
\begin{equation} \label{eq:DiracKillingForm}
\killing{P_{\alpha}}{P_{\beta}}_{\mathfrak{l}} = 0\,, \qquad
\killing{P^{\alpha}}{P^{\beta}}_{\mathfrak{l}} = 0\,, \qquad
\killing{P_{\alpha}}{P^{\beta}}_{\mathfrak{l}} = - \delta^{\beta}_{\alpha}\,, \qquad
\killing{P^{\alpha}}{P_{\beta}}_{\mathfrak{l}} = \delta_{\beta}^{\alpha}\,.
\end{equation}
Note that, for general $A,B\in\mathfrak{l}$, we can write
\begin{equation}
\killing{A}{B}_{\mathfrak{l}} = A^I B^J g_{IJ}\,, \qquad \killing{B}{A}_{\mathfrak{l}} = B^J A^I g_{JI}\,.
\end{equation}
Since we require the bilinear form to be symmetric, the two expressions must be equal. However, since the coefficients in $\mathfrak{l}$ are Grassmann numbers, we have $A^IB^J= -B^JA^I$, so it follows that $g_{IJ} = - g_{JI}$. Hence the antisymmetry of (\ref{eq:DiracKillingForm}).

Now we use the properties of the group $L$ and the action $\rhd$ of $G$ on $L$ to recognize the physical nature of the Lagrange multiplier $D$ in (\ref{eq:bfcgdh}). Indeed, the choice of the group $L$ dictates that $D$ contains $8$ independent complex Grassmannian matter fields as its components. Moreover, due to the fact that $D$ is a $0$-form and that it transforms according to the spinorial representation of $SO(3,1)$, we can identify its components with the Dirac bispinor fields, and write
\begin{equation}
D = \psi^{\alpha} P_{\alpha} + \bar{\psi}_{\alpha} P^{\alpha}\,,
\end{equation}
where it is assumed that $\psi$ and $\bar\psi$ are independent fields, as usual. This is again an illustration of the fact that information about the structure of the matter sector in the theory is specified by the choice of the group $L$ in the $2$-crossed module, and another main result of the paper.

Given all of the above, now we can finally write the $3BF$ action (\ref{eq:bfcgdh}) corresponding to this choice of the $2$-crossed module as
\begin{equation}\label{eq:DiracTopoloski}
    S_{3BF}=\int_{\cM_4} B^{ab}\wedge R_{ab} + e_a\wedge \nabla \beta^a + (\bar{\gamma}{\overset{\leftarrow}{\nabla}) {}_{\alpha} \psi^{\alpha} +\bar{\psi}_{\alpha}{({\overset{\rightarrow}{\nabla}}\gamma)}{}^{\alpha}}\,.
\end{equation}
In order to promote this action into a full theory of gravity coupled to Dirac fermions, we add the convenient constraint terms to the action, as follows:
\begin{equation}\label{eq:Dirac}
\begin{aligned}
S=\int_{\cM_4} & B^{ab}\wedge R_{ab} + e_a\wedge \nabla \beta^a + ( \bar{\gamma} {\overset{\leftarrow}{\nabla}} ) {}_{\alpha} \psi^{\alpha} +\bar{\psi}_{\alpha}{({\overset{\rightarrow}{\nabla}}\gamma)}{}^{\alpha} \\
&- \lambda_{ab} \wedge \Big(B^{ab}\vphantom{\ds\int}-\frac{1}{16\pi l_p^2}\varepsilon^{abcd} e_c \wedge e_d\Big)\vphantom{\ds\int}\\
 &-\lambda^{\alpha}\wedge \Big({\bar{\gamma}}_{\alpha}-\frac{i}{6}\varepsilon_{abcd}e^a\wedge e^b \wedge e^c (\bar{\psi}\gamma^d)_{\alpha}\Big)+\bar{\lambda}_{\alpha}\wedge \Big({\gamma}^{\alpha}+\frac{i}{6}\varepsilon_{abcd}e^a\wedge e^b \wedge e^c (\gamma^d\psi){}^{\alpha}\Big)\vphantom{\ds\int}\\
 & -\frac{1}{12} m \, \bar{\psi}\psi\, \varepsilon_{abcd}e^a\wedge e^b \wedge e^c \wedge e^d +2 \pi i l_p^2 \, \bar{\psi}\gamma_5\gamma^a \psi \, \varepsilon_{abcd}e^b\wedge e^c \wedge \beta^d \,.\vphantom{\ds\int}
\end{aligned}
\end{equation}
Here the first row is the topological sector, the second row is the gravitational simplicity constraint term from (\ref{eq:GravityVeza}), while the third row contains the new simplicity constraints for the Dirac field corresponding to the Lagrange multiplier $1$-forms $\lambda^{\alpha}$ and $\bar{\lambda}_{\alpha}$. The fourth row contains the mass term for the Dirac field, and a term which ensures the correct coupling between the torsion and the spin of the Dirac field, as specified by the Einstein-Cartan theory. Namely, we want to ensure that the torsion has the form
\begin{equation} \label{eq:DiracTorzija}
T_a \equiv \nabla e_a = 2\pi l_p^2 s_a\,,
\end{equation}
where
\begin{equation}\label{eq:torzija}
s_a = i\varepsilon_{abcd} e^b \wedge e^c \bar\psi \gamma_5 \gamma^d \psi
\end{equation}
is the spin $2$-form. Of course, other couplings should also be straightforward to implement, but we choose this particular coupling because we are interested in reproducing the standard Einstein-Cartan gravity coupled to the Dirac field.

Varying the action (\ref{eq:Dirac}) with respect to $B_{ab}$, $\lambda^{ab}$, $\bar{\gamma}_{\alpha}$, $\gamma^{\alpha}$, $\lambda^{\alpha}$, ${\bar{\lambda}}_{\alpha}$, $\bar{\psi}{}_{\alpha}$, $\psi^{\alpha}$, $e^a$, $\beta^a$ and $\omega^{ab}$ one obtains the equations of motion:
\begin{gather}
\label{eq:d00}
R^{ab}-\lambda^{ab}=0\,,\vphantom{\ds\int}\\
\label{eq:d000}
B_{ab}-\frac{1}{16\pi l_p^2}\varepsilon_{abcd} e^c \wedge e^d=0\,,\vphantom{\ds\int}\\
\label{eq:d1}
(\overset{\rightarrow}{\nabla}\psi)^{\alpha}-\lambda^ {\alpha}=0\,,\vphantom{\ds\int}\\
\label{eq:d2}
(\bar{\psi}\overset{\leftarrow}{\nabla})_{\alpha}-{\bar{\lambda}}_{\alpha}=0\,,\vphantom{\ds\int}\\
\label{eq:d3}
\bar{\gamma}{}_{\alpha}-\frac{i}{6}\varepsilon_{abcd}e^a\wedge e^b \wedge e^c (\bar{\psi}\gamma^d)_{\alpha}=0\,,\vphantom{\ds\int}\\
\label{eq:d4}
{\gamma}^{\alpha}+\frac{i}{6}\varepsilon_{abcd}e^a\wedge e^b \wedge e^c (\gamma^d\psi)^{\alpha}=0\,,\vphantom{\ds\int}\\
\label{eq:d5}
\begin{aligned}
\D {\gamma}^{\alpha}+{\omega^{\alpha}}_{\beta}\wedge{\gamma}^{\beta}&+\frac{i}{6}\lambda^{\beta}\wedge\varepsilon_{abcd}e^a\wedge e^b \wedge e^c {{\gamma^d}{}^{\alpha}}_{\beta}+\frac{1}{12}m\varepsilon_{abcd}e^a\wedge e^b \wedge e^c \wedge e^d \psi^{\alpha}\vphantom{\ds\int}\\&+i2\pi l_p^2\varepsilon_{abcd}e^a\wedge e^b \wedge \beta^c (\gamma_5\gamma^d\psi)^{\alpha}=0\,,\vphantom{\ds\int}
\end{aligned}\\
\label{eq:d5.01}
\begin{aligned}
\D \bar{{\gamma}}{}_{\alpha}-\bar{\gamma}{}_{\beta}\wedge{\omega^{\beta}}{}_{\alpha}&+\frac{i}{6}\bar{\lambda}{}_{\beta}\wedge\varepsilon_{abcd}e^a\wedge e^b \wedge e^c {{\gamma^d}{}^{\beta}}_{\alpha}-\frac{1}{12}m\varepsilon_{abcd}e^a\wedge e^b \wedge e^c \wedge e^d \bar{\psi}{}_{\alpha}\vphantom{\ds\int}\\&-i2\pi l_p^2\varepsilon_{abcd}e^a\wedge e^b \wedge \beta^c (\bar{\psi}\gamma_5\gamma^d)_{\alpha}=0\,,\vphantom{\ds\int}
\end{aligned}\\
\label{eq:d6}
\begin{aligned}
   \nabla \beta_a + 2 \varepsilon_{abcd} & \lambda^{bc} \wedge e^d-\frac{i}{2}\varepsilon_{abcd}\lambda^{\alpha}\wedge e^b\wedge e^c (\bar{\psi}\gamma^d)_{\alpha}+\frac{i}{2}\varepsilon_{abcd}{\bar{\lambda}}_{\alpha}\wedge e^b\wedge e^c (\gamma^d\psi)^{\alpha} \vphantom{\ds\int}\\& -\frac{1}{3}\varepsilon_{abcd} e^b \wedge e^c \wedge e^d m\bar{\psi}\psi -4\pi l_p^2i\varepsilon_{abcd}e^b\wedge \beta^c \bar{\psi}\gamma_5 \gamma^d \psi=0\,,\vphantom{\ds\int}
\end{aligned}\\
\label{eq:d7}
    \nabla e_a-i2\pi l_p^2\varepsilon_{abcd}e^b\wedge e^c \bar{\psi}\gamma_5\gamma^d \psi=0\,,\vphantom{\ds\int}\\
\label{eq:d8}
    \nabla B_{ab} - e_{[a} \wedge \beta_{b]}+\bar{\gamma}\frac{1}{8}[\gamma_a,\gamma_b]\psi+\bar{\psi}\frac{1}{8}[\gamma_a,\gamma_b]\gamma=0\,.\vphantom{\ds\int}
\end{gather}
The dynamical degrees of freedom are $e^a$, $\psi^{\alpha}$ and $\bar{\psi}_{\alpha}$, while the remaining variables are determined in terms of the dynamical variables, and are given as:
\begin{equation}
    \begin{array}{c}\label{eq:d9}
    B_{ab}{}_{\mu\nu}=\frac{1}{8\pi l_p^2}\varepsilon_{abcd}e^c{}_\mu e^d{}_\nu\,, \qquad  \lambda^ {\alpha}{}_\mu=(\overset{\rightarrow}{\nabla}_\mu\psi)^{\alpha}\,, \qquad   {\bar{\lambda}}_{\alpha}{}_\mu=(\bar{\psi}\overset{\leftarrow}{\nabla}_\mu)_{\alpha}\,, \vphantom{\ds\int}\\
    \bar{\gamma}{}_{\alpha}{}_{\mu\nu\rho}=i\varepsilon_{abcd}e^a{}_\mu e^b{}_\nu  e^c{}_\rho(\bar{\psi}\gamma^d)_{\alpha}\,, \qquad
    {\gamma}^{\alpha}{}_{\mu\nu\rho}=-i\varepsilon_{abcd}e^a{}_\mu e^b{}_\nu  e^c{}_\rho (\gamma^d\psi)^{\alpha}\,, \vphantom{\ds\int} \\
\lambda_{ab}{}_{\mu\nu}=R_{ab}{}_{\mu\nu}\,, \qquad \omega^{ab}{}_\mu=\triangle^{ab}{}_\mu+K^{ab}{}_\mu\,. \vphantom{\ds\int} \\
\end{array}
\end{equation}
Here $K^{ab}{}_{\mu}$ is the contorsion tensor, constructed in the standard way from the torsion tensor, whereas from (\ref{eq:d7}) we have
\begin{equation} \label{eq:DiracTorzijaJosJednom}
T_a \equiv \nabla e_a = 2\pi l_p^2 s_a\,,
\end{equation}
which is precisely the desired equation (\ref{eq:DiracTorzija}). Further, from the equation (\ref{eq:d000}) one obtains
\begin{equation}
\nabla B^{ab}=-\frac{1}{8\pi l_p^2}\varepsilon^{abcd} \left(e_c \wedge \nabla e_d\right)\,.
\end{equation}
Substituting this expression in the equation (\ref{eq:d8}) it follows that
\begin{equation}
    2\varepsilon_{abcd} e^c \wedge\left(- \frac{1}{16\pi l_p^2} \nabla e^d+\frac{1}{8} s^d\right)-e_{[a}\wedge \beta_{b]}=0\,. 
\end{equation}
The expression in the parentheses is equal to zero, according to the equation (\ref{eq:d7}). From the remaining term $e_{[a}\wedge \beta_{b]}=0$ it again follows that
\begin{equation} \label{eq:360}
\beta=0\,.
\end{equation}
Using this result, the equation of motion (\ref{eq:d5}) for fermions becomes
\begin{equation}
   \frac{i}{6} \varepsilon_{abcd} e^a\wedge e^b \wedge \left( 2e^c \wedge \gamma^d \overset{\rightarrow}{\nabla} + \frac{im}{2} e^c\wedge e^d
 - 3 ( \nabla e^c ) \gamma^d \right)\psi = 0\,.
\end{equation}
Using equation (\ref{eq:d7}), the last term in the parentheses vanishes, and the equation reduces to the covariant Dirac equation,
\begin{equation} \label{eq:DiracEquation}
    (i\gamma^a e^{\mu}{}_a \overset{\rightarrow}{\nabla}_\mu -m)\psi=0\,,
\end{equation}
where $e^{\mu}{}_a$ is the inverse tetrad. Similarly, the equation (\ref{eq:d5.01}) gives the conjugated Dirac equation:
\begin{equation} \label{eq:CdiracEquation}
    \bar{\psi}(i\overset{\leftarrow}{\nabla}_\mu e^{\mu}{}_a \gamma^a+m)=0\,.
\end{equation}
Finally, the equation of motion (\ref{eq:d6}) for tetrad field reduces to
\begin{equation}\label{eq:diraceomfore}
    R^{\mu\nu}-\frac{1}{2}g^{\mu\nu}R=8\pi l_p^2\; T^{\mu\nu}\,,\quad \quad T^{\mu\nu} \equiv \frac{i}{2}\bar{\psi} \gamma^\nu{\overset{\leftrightarrow}{\nabla}}{}^a e^\mu{}_a \psi-\frac{1}{2}g^{\mu\nu}\bar{\psi} \Big(i\gamma^a\overset{\leftrightarrow}{\nabla}_\rho e^\rho{}_a-2m \Big)\psi\,,
\end{equation}
Here, we used the notation ${\overset{\leftrightarrow}{\nabla}}={\overset{\rightarrow}{\nabla}}-{\overset{\leftarrow}{\nabla}}$. The system of equations (\ref{eq:d00})-(\ref{eq:d8}) is equivalent to the system of equations (\ref{eq:d9}), (\ref{eq:360}), (\ref{eq:DiracEquation})-(\ref{eq:diraceomfore}). As we expected, the equations of motion (\ref{eq:DiracTorzijaJosJednom}), (\ref{eq:DiracEquation}), (\ref{eq:CdiracEquation}) and (\ref{eq:diraceomfore}) are precisely the equations of motion of the Einstein-Cartan theory coupled to a Dirac field.

\subsection{Constrained $3BF$ action for the Weyl and Majorana fields}\label{sec:III.4}

A general solution of the Dirac equation is not an irreducible representation of the Lorentz group, and one can rewrite Dirac fermions as left-chiral and right-chiral fermion fields that both retain their chirality under Lorentz transformations, implying their irreducibility. Hence, it is useful to rewrite the action for left and right Weyl spinors as a constrained $3BF$ action. For simplicity, we will discuss only left-chiral spinor field, while the right-chiral field can be treated analogously. Both Weyl and Majorana fermions can be treated in the same way, the only difference being the presence of an additional mass term in the Majorana action.

We being by defining a $2$-crossed module $(L\stackrel{\delta}{\to} H \stackrel{\partial}{\to}G\,, \rhd\,, \{\_\,,\_\})$, as follows. The groups are:
\begin{equation}
G=SO(3,1)\,, \quad \quad H=\mathbb{R}^4\,, \quad \quad L=\mathbb{R}^4(\grasmanovi)\,.
\end{equation}
The maps $\partial$,  $\delta$ and the Peiffer lifting are trivial. The action $\rhd$ of the group $G$ on $G$, $H$ and $L$ is given in the same way as for the Dirac case, whereas the spinorial representation reduces to
\begin{equation} \label{eq:actionOfGonLweyl}
M_{ab}\rhd P^{\alpha}=\frac{1}{2}(\sigma_{ab}){}^{\alpha}{}_{\beta} P^{\beta}\,, \qquad M_{ab} \rhd P_{\dot{\alpha}}=\frac{1}{2}(\bar{\sigma}{}_{ab}){}^{\dot{\beta}}{}_{\dot{\alpha}} P_{\dot{\beta}}\,,
\end{equation}
where $\sigma^{ab}=-\bar{\sigma}{}^{ab}=\frac{1}{4}(\sigma^a\bar{\sigma}^b-\sigma^b\bar{\sigma}^a)$, for $\sigma^a = (1, \vec{\sigma})$ and $\bar{\sigma}{}^a=(1,-\vec{\sigma})$, in which $\vec{\sigma}$ denotes the set of three Pauli matrices. The four generators of the group $L$ are denoted as $P^{\alpha}$ and $P_{\dot{\alpha}}$, where the Weyl indices $\alpha,\dot{\alpha}$ take values $1,2$.

The $3$-connection $(\alpha\,,\beta\,,\gamma)$ now takes the form corresponding to this choice of Lie groups,
\begin{equation}
\alpha = \omega^{ab}M_{ab}\,, \quad \quad \beta=\beta^a P_a\,, \quad \quad \gamma = \gamma_{\alpha} P^{\alpha}+ \bar{\gamma}{}^{\dot{\alpha}}P_{\dot{\alpha}}\,,
\end{equation}
while the fake $3$-curvature $(\cF\,, \cG\,, \cH)$ defined in (\ref{eq:3krivine}) is
\begin{equation}
\begin{aligned}
     \cF = R^{ab} M_{ab}\,, \quad & \quad \cG= \nabla \beta^a P_a\,,\\
     \quad \cH= \big(\D \gamma_{{\alpha}}+\frac{1}{2}\omega^{ab}(\sigma^{ab}){}^{{\beta}}{}_{\alpha}\gamma_{\beta}\big)P^{\alpha}+\big( \D {\bar{\gamma}}{}^{\dot{\alpha}}&+\frac{1}{2}\omega_{ab}({\bar{\sigma}}{}^{ab}){}^{\dot{\alpha}}{}_{\dot{\beta}}\bar{\gamma}^{\dot{\beta}}  \big) P{}_{\dot{\alpha}}  \equiv (\overset{\rightarrow}{\nabla} \gamma){}_{\alpha} P^{\alpha} + ({\bar{\gamma}}\overset{\leftarrow}{\nabla}){}^{\dot{\alpha}}P{}_{\dot{\alpha}}\,.
\end{aligned}
\end{equation}
Introducing the spinor fields $\psi_{\alpha}$ and $\bar{\psi}^{\dot{\alpha}}$ via the Lagrange multiplier $D$ as
\begin{equation}
D = \psi_{\alpha} P^{\alpha} + \bar{\psi}^{\dot{\alpha}} P_{\dot{\alpha}}\,,
\end{equation}
and using the bilinear form $\killing{\_}{\_}_{\mathfrak{l}}$ for the group $L$,
\begin{equation}
\killing{P^{\alpha}}{P^{\beta}}_{\mathfrak{l}} = \lc^{\alpha\beta}\,, \qquad
\killing{P_{\dot{\alpha}}}{P_{\dot{\beta}}}_{\mathfrak{l}} = \lc_{\dot{\alpha}\dot{\beta}}\,, \qquad
\killing{P^{\alpha}}{P_{\dot{\beta}}}_{\mathfrak{l}} = 0 \,, \qquad
\killing{P_{\dot{\alpha}}}{P^{\beta}}_{\mathfrak{l}} = 0 \,,
\end{equation}
where $\lc^{\alpha\beta}$ and $\lc_{\dot{\alpha}\dot{\beta}}$ are the usual two-dimensional antisymmetric Levi-Civita symbols, the topological $3BF$ action (\ref{eq:bfcgdh}) for spinors coupled to gravity becomes
\begin{equation}\label{eq:Weyltopoloski}
    S_{3BF}=\int_{\cM_4} B^{ab}\wedge R_{ab}+e_a\wedge\nabla\beta^a+ \psi{}^\alpha\wedge (\overset{\rightarrow}{\nabla} \gamma)_\alpha+ \bar{\psi}{}{}_{\dot{\alpha}}\wedge (\bar{\gamma} \overset{\leftarrow}{\nabla}){}^{\dot{\alpha}}\,.
\end{equation}
In order to obtain the suitable equations of motion for the Weyl spinors, we again introduce appropriate simplicity constraints, so that the action becomes:
\begin{equation}\label{eq:Weyl}
\begin{aligned}
    S=&\int_{\cM_4} B^{ab}\wedge R_{ab}+e_a\wedge\nabla\beta^a+\psi{}^\alpha\wedge (\overset{\rightarrow}{\nabla} \gamma)_\alpha+ \bar{\psi}{}{}_{\dot{\alpha}}\wedge (\bar{\gamma} \overset{\leftarrow}{\nabla}){}^{\dot{\alpha}}\\&- \lambda_{ab} \wedge (B^{ab}-\frac{1}{16\pi l_p^2}\varepsilon^{abcd} e_c \wedge e_d)\vphantom{\ds\int}\\&-\lambda{}^\alpha\wedge ( \gamma{}_\alpha+\frac{i}{6} \varepsilon_{abcd}e^a\wedge e^b \wedge e^c\sigma^d{}_{\alpha\dot{\beta}}\bar{\psi}{}^{\dot{\beta}})-\bar{\lambda}{}_{\dot{\alpha}}\wedge(\bar{\gamma}{}^{\dot{\alpha}}+\frac{i}{6} \varepsilon_{abcd}e^a\wedge e^b \wedge e^c \bar{\sigma}{}^d{}^{\dot{\alpha}\beta}\psi_\beta)\vphantom{\ds\int}\\&-4\pi l_p^2 \varepsilon_{abcd} e^a\wedge e^b \wedge \beta^c (\bar{\psi}{}_{\dot{\alpha}}\bar{\sigma}{}^d{}^{\dot{\alpha}\beta}\psi_\beta)\,.\vphantom{\ds\int}
\end{aligned}
\end{equation}
The new simplicity constraints are in the third row, featuring the Lagrange multiplier $1$-forms $\lambda{}_\alpha$ and $\bar{\lambda}{}^{\dot{\alpha}}$. Also, using the coupling between the Dirac field and torsion from Einstein-Cartan theory as a model, the term in the fourth row is chosen to ensure that the coupling between the Weyl spin tensor
\begin{equation}
s_a \equiv i\varepsilon_{abcd} e^b \wedge e^c \;\psi^\alpha \sigma^d{}_{\alpha\dot{\beta}}\bar{\psi}{}^{\dot{\beta}}\,,
\end{equation}
and torsion is given as:
\begin{equation}
T_a=4\pi l_p^2 s_a\,.
\end{equation}
The case of the Majorana field is introduced in exactly the same way, albeit with an additional mass term in the action, of the form:
\begin{equation} \label{eq:MajoranaMassTerm}
-\frac{1}{12} m \varepsilon_{abcd}e^a\wedge e^b \wedge e^c  \wedge e^d ( \psi^\alpha  \psi_\alpha +\bar{\psi}_{\dot{\alpha}}\bar{\psi}^{\dot{\alpha}} ) \,.
\end{equation}

Varying the action (\ref{eq:Weyl}) with respect to the variables $B_{ab}$, $\lambda^{ab}$, $\gamma{}_\alpha$, $\bar{\gamma}{}^{\dot{\alpha}}$, $\lambda{}_\alpha$, $\bar{\lambda}{}{}^{\dot{\alpha}}$, $\psi_\alpha$, $\bar{\psi}{}^{\dot{\alpha}}$, $e^a$, $\beta^a$ and $\omega^{ab}$ one again obtains the complete set of equations of motion, displayed in the Appendix \ref{ApendiksC}. The only dynamical degrees of freedom are $\psi_\alpha$, $\bar{\psi}{}^{\dot{\alpha}}$ and $e^a$, while the remaining variables are algebraically determined in terms of these as:
\begin{equation} \label{eq:WeylLagMult}
    \begin{aligned}
&\lambda^{ab}{}_{\mu\nu}=R^{ab}{}_{\mu\nu}\,, \quad B_{ab}{}_{\mu\nu}=\frac{1}{8\pi l_p^2}\varepsilon_{abcd} e^c{}_{\mu} e^d{}_{\nu}\,, \quad 
\lambda{}_\alpha{}_\mu=\nabla_\mu\psi_\alpha\,, \quad 
\bar{\lambda}{}^{\dot{\alpha}}{}_\mu=\nabla_\mu \bar{\psi}{}^{\dot{\alpha}}\,,\vphantom{\ds\int}\\
\gamma{}_\alpha{}_{\mu\nu\rho}&=i\varepsilon_{abcd}e^a{}_\mu e^b{}_\nu e^c{}_\rho \sigma^d{}_{\alpha\dot{\beta}}\bar{\psi}{}^{\dot{\beta}}\,,\quad
\bar{\gamma}{}^{\dot{\alpha}}{}_{\mu\nu\rho}=i\varepsilon_{abcd}e^a{}_\mu e^b{}_\nu e^c{}_\rho \bar{\sigma}{}^d{}^{\dot{\alpha}\beta}\psi_\beta\,, \quad \omega_{ab\mu}=\triangle_{ab\mu}+K_{ab\mu}\,.\vphantom{\ds\int}
\end{aligned}
\end{equation}
In addition, one also maintains the result $\beta=0$ as before. Finally, the equations of motion for the dynamical fields are
\begin{equation}
\bar{\sigma}{}^{a\dot{\alpha}\beta} e^{\mu}{}_a \nabla_\mu \psi_\beta=0\,, \qquad
\sigma^a{}_{\alpha\dot{\beta}} e^{\mu}{}_a \nabla_\mu \bar{\psi}^{\dot{\beta}}=0\,,
\end{equation}
and
\begin{equation}
R^{\mu\nu}-\frac{1}{2}g^{\mu\nu}R=8\pi l_p^2 \; T^{\mu\nu}\,,
\end{equation}
where
\begin{equation}
T^{\mu\nu} \equiv \frac{i}{2}\bar{\psi}\bar{\sigma}{}^b e^{\nu}{}_b \nabla^\mu\psi+\frac{i}{2}\psi\sigma^b e^{\nu}{}_b \nabla^\mu\bar{\psi}-g^{\mu\nu}\frac{1}{2}\Big(i\bar{\psi}\bar{\sigma}^a e^{\lambda}{}_a \nabla_{\lambda} \psi+i\psi\sigma^a e^{\lambda}{}_a \nabla_{\lambda}\bar{\psi}\Big)\,.
\end{equation}
Here we have suppressed the spinor indices. In the case of the Majorana field, the equations of motion (\ref{eq:WeylLagMult}) remain the same, while the equations of motion for $\psi_\alpha$ and $\bar{\psi}{}^{\dot{\alpha}}$ take the form
\begin{equation}\label{eq:m12}
   i \sigma^a{}_{\alpha\dot{\beta}} e^{\mu}{}_a\nabla_\mu \bar{\psi}{}^{\dot{\beta}} -m\psi_\alpha =0\,, \qquad
  i \bar{\sigma}{}^{a\dot{\alpha}\beta} e^{\mu}{}_a \nabla_\mu \psi_\beta - m\bar{\psi}{}^{\dot{\alpha}} =0\,,
\end{equation}
whereas the stress-energy tensor takes the form
\begin{equation}
\begin{array}{lcl}
T^{\mu\nu} & \equiv & \ds \frac{i}{2}\bar{\psi}\bar{\sigma}{}^b e^{\nu}{}_b \nabla^\mu\psi+\frac{i}{2}\psi\sigma^b e^{\nu}{}_b \nabla^\mu\bar{\psi} \vphantom{\ds\int} \\
 & & \ds - g^{\mu\nu}\frac{1}{2}\left[i\bar{\psi}\bar{\sigma}^a e^{\lambda}{}_a \nabla_{\lambda} \psi+i\psi\sigma^a e^{\lambda}{}_a \nabla_{\lambda}\bar{\psi}-\frac{1}{2}m\left(\psi\psi+\bar{\psi}\bar{\psi}\right)\right]\,. \\
\end{array}
\end{equation}

\bigskip




\section{\label{SecIV}The Standard Model}


The Standard Model $3$-group can be defined as: 
\begin{equation}
    G = SO(3,1) \times SU(3) \times SU(2) \times U(1)\,, \quad H=\mathbb{R}^4\,, \quad  L= \mathbb{R}^4(\mathbb{C}) \times \mathbb{R}^{64}(\grasmanovi) \times \mathbb{R}^{64}(\grasmanovi) \times \mathbb{R}^{64}(\grasmanovi)\,,
\end{equation}
where $\mathbb{C}$ denotes the field of complex numbers. The motivation for this choice of the group $L$ is given in the table below.
\begin{center}
\setlength{\tabcolsep}{3pt}
    \label{tab:table3}
\begin{tabular}{|c|c|c|c|} \hline
1. lepton generation &\shortstack{red color \strut \\ 1. quark generation} & \shortstack{green color \strut \\ 1. quark generation} & \shortstack{blue color \strut \\ 1. quark generation} \\
\hline \hline
 \rule{0pt}{0.7cm}
\Vcentre{$\begin{pmatrix}
\nu_e \\
e^-
\end{pmatrix}_L$}{} \rule[-0.7cm]{0pt}{0.7cm} & \Vcentre{$\begin{pmatrix}
u_r \\ d_r \end{pmatrix}_L$} & \Vcentre{$\begin{pmatrix}
u_g \\ d_g
\end{pmatrix}_L$} & \Vcentre{$\begin{pmatrix}
u_b \\ d_b
\end{pmatrix}_L$}$\vphantom{\ds\int\frac{A^A}{B}} $\\[12pt] \hline
$ \vphantom{\ds\int} (\nu_e){}_R$ &  $(u_r){}_R$ & $(u_g){}_R$ & $(u_b){}_R $ \\ \hline
$(e^-){}_R \vphantom{\ds\int}$  & $(d_r){}_R$ & $(d_g){}_R$ & $(d_b){}_R$ \\
\hline

\end{tabular}
\end{center}

We see that in order to introduce one generation of matter one needs to provide $16$ spinors, or equivalently the group $L$ has to be chosen as $L=\mathbb{R}^{64}(\grasmanovi)$. As there are three generations of matter, the part of the group $L$ that corresponds to the fermion fields in the theory is chosen to be $L=\mathbb{R}^{64}(\grasmanovi)\times\mathbb{R}^{64}(\grasmanovi)\times\mathbb{R}^{64}(\grasmanovi)$. To define the Higgs sector one needs two complex scalar fields $\begin{pmatrix} \phi^+ \\ \phi_0 \end{pmatrix}$, or equivalently the scalar sector of the group $L$ is given as $L=\mathbb{R}^4(\mathbb{C})$.

The maps $\partial$,  $\delta$ and the Peiffer lifting are trivial. The action of the group $G$ on itself is given via conjugation. The action of the $SO(3,1)$ subgroup of $G$ on $H$ is via 
vector representation and the action of $SU(3) \times SU(2) \times U(1)$ subgroup on $H$ is via trivial representation. The action of the $SO(3,1)$ on $L$ is via trivial representation for the generators corresponding to the scalar fields, i.e. the $\mathbb{R}^4(\mathbb{C})$ subgroup of $L$, and via spinor representation for the every quadruple of generators corresponding to the fermion fields, given as in the section \ref{SecIII}. The information how spinors transform under the $SU(3)\times SU(2)\times U(1)$ group is encoded in the action of that subgroup of $G$ on $L$, as specified in the table above. For simplicity, in the following, only one family of the lepton sector and only electroweak part of the gauge sector of the Standard model is considered.

The groups are chosen as:
\begin{equation}
    G = SO(3,1) \times SU(2) \times U(1)\,, \quad H=\mathbb{R}^4\,, \quad  L^{\rm leptons}=  \mathbb{R}^{16}(\grasmanovi) \times \mathbb{R}^4(\mathbb{C}) \,.
\end{equation}
The $3$-connection then takes the form
\begin{equation}
\begin{array}{c}
    \alpha=\omega^{ab}M_{ab}+W^IT_I+AY\,, \qquad \beta=\beta^aP_a\,, \vphantom{\ds\int} \\ 
\ds \gamma= \gamma_\alpha{}^{\tilde{L}} P^\alpha{}_{\tilde{L}} + \gamma^{\dot{\alpha}}{}_{\tilde{L}} P_{\dot{\alpha}}{}^{\tilde{L}}+\gamma_\alpha{}^{\tilde{R}} P^\alpha{}_{\tilde{R}} + \gamma^{\dot{\alpha}}{}_{\tilde{R}} P_{\dot{\alpha}}{}^{\tilde{R}}+\gamma^{\tilde{a}}P_{\tilde{a}}\,.  \vphantom{\ds\int} \\
\end{array}
\end{equation}
Here the indices $I,J,...$ take the values $1,2,3$ and counts the Pauli matrices, generators of the group $SU(2)$, the indices $\tilde{L},\tilde{L}',...$ take the values $1,2$ and count the components of left doublet, $\tilde{R}$ denotes the right singlet $(e{}^-)_R$ and right singlet $(\nu_e)_R$, and indices ${\tilde{a}},\tilde{b},..$ take values $1,2$ and count the components of the scalar doublet. It is also useful to define $\tilde{i}=(\tilde{L},\tilde{R})$ which takes values $1,\dots,4$.

The action of the group $G$ on $L$ is defined as:
\begin{equation}
\begin{array}{c}
\ds    M_{ab}\rhd P^{\alpha}{}_i=\frac{1}{2}(\sigma_{ab}){}^{\alpha}{}_{\beta} P^{\beta}{}_i\,, \qquad M_{ab} \rhd P_{\dot{\alpha}}{}_i=\frac{1}{2}(\bar{\sigma}{}_{ab}){}^{\dot{\beta}}{}_{\dot{\alpha}} P_{\dot{\beta}i}\,, \qquad M_{ab}\rhd P_{\tilde{a}}=0 \,,\vphantom{\ds\int}\\
\ds T_I\rhd P^\alpha{}_{\tilde{L}}=\frac{1}{2}(\sigma_I){}^{\tilde{L}'}{}_{\tilde{L}} P^\alpha{}_{\tilde{L'}}\,, \qquad T_I\rhd P_{\dot{\alpha}}{}_{\tilde{L}}=\frac{1}{2}(\sigma_I){}^{\tilde{L}'}{}_{\tilde{L}} P_{\dot{\alpha}}{}_{\tilde{L}'}\,, \vphantom{\ds\int} \\
\ds T_I\rhd P^\alpha{}_{\tilde{R}}=0\,,\qquad T_I\rhd P_{\dot{\alpha}}{}_{\tilde{R}}=0\,, \qquad T_I\rhd P_{\tilde{a}} = \frac{1}{2}(\sigma_I){}^{\tilde{b}}{}_{\tilde{a}} P_{\tilde{b}} \,,\vphantom{\ds\int}\\
    Y\rhd P^\alpha{}_{\tilde{L}}=-P^\alpha{}_{\tilde{L}}\,, \quad Y\rhd P^\alpha{}_{e_R} =-2P^\alpha{}_{e_R}\,,\quad Y\rhd P^\alpha{}_{\nu_R} =-2P^\alpha{}_{\nu_R}\,, \quad Y\rhd P_{\tilde{a}} = P_{\tilde{a}} \,,\vphantom{\ds\int}\\
    Y\rhd P_{\dot{\alpha}}{}_{\tilde{L}}=-P_{\dot{\alpha}}{}_{\tilde{L}}\,, \qquad Y\rhd P_{\dot{\alpha}}{}_{e_R}=-2P_{\dot{\alpha}}{}_{e_R}\,, \qquad Y\rhd P_{\dot{\alpha}}{}_{\nu_R}=-2P_{\dot{\alpha}}{}_{\nu_R}\,.\vphantom{\ds\int} \\
\end{array}
\end{equation}

The $3$-curvatures are given as:
\begin{equation}
    \begin{aligned}
     \cF=R^{ab}M_{ab}+F^IT_I + F Y\,, \qquad& \cG= \nabla \beta^a P_a\,,\\
     \cH =  (\overset{\rightarrow}{\nabla} \gamma{}^{\tilde{L}}){}_{\alpha} P^{\alpha}{}_{\tilde{L}} + ({\bar{\gamma}{}_{\tilde{L}}}\overset{\leftarrow}{\nabla}){}^{\dot{\alpha}}P{}_{\dot{\alpha}}{}^{\tilde{L}}+  (\overset{\rightarrow}{\nabla} \gamma{}^{\tilde{R}}){}_{\alpha} &P^{\alpha}{}_{\tilde{R}} + ({\bar{\gamma}{}_{\tilde{R}}}\overset{\leftarrow}{\nabla}){}^{\dot{\alpha}}P{}_{\dot{\alpha}}{}^{\tilde{R}} +  \D \gamma^{\tilde{a}}P_{\tilde{a}}\,.
    \end{aligned}
\end{equation} 
The topological $3BF$ action is defined as:
\begin{equation}
    S=\int B_{ab}R^{ab}+B_I F^I + BF + e_a\nabla \beta^a +  \psi^{\alpha}{}_{\tilde{i}}(\overset{\rightarrow}{\nabla} \gamma{}^{\tilde{i}}){}_{\alpha} + \bar{\psi}_{\dot{\alpha}}{}^{\tilde{i}}({\bar{\gamma}{}_{\tilde{i}}}\overset{\leftarrow}{\nabla}){}^{\dot{\alpha}} +   \phi^{\tilde{a}} \D \gamma_{\tilde{a}}\,.
\end{equation}
At this point, it is useful to simplify the notation and denote all indices of the group $G$ by $\hat{\alpha}$, of the group $H$ by $\hat{a}$ and $L$ by $\hat{A}$. In order to promote this action to a full theory of first lepton family coupled to electroweak gauge fields, Higgs field, and gravity, we again introduce the appropriate simplicity constraint, as follows
\begin{equation}
\begin{aligned}
   S =\int  &B_{\hat{\alpha}}\wedge \mathcal{F}^{\hat{\alpha}}+e_{\hat {a}}\wedge \mathcal{G}^{\hat{a}}+D_{\hat{A}}\wedge \mathcal{H}^{\hat{A}}\vphantom{\ds\int}\\ &+ \left(B_{\hat{\alpha}}-C_{\hat{\alpha}}{}^{\hat{\beta}}M_{cd\hat{\beta}} e^c\wedge e^d\right)\wedge\lambda^{\hat{\alpha}}-  \left(\gamma_{\hat{A}}-e^a\wedge e^b \wedge e^c C_{\hat{A}}{}^{\hat{B}}M_{abc\hat{B}}\right)\wedge{\lambda}^{\hat{A}}\vphantom{\ds\int}\\ &+{\zeta^{ab}{}_{\hat{\alpha}}}\wedge\left({M{}_{ab}{}^{\hat{\alpha}}}\varepsilon^{cdef}e_c\wedge e_d \wedge e_e \wedge e_f - F^{\hat{\alpha}} \wedge e_c \wedge e_d\right)\vphantom{\ds\int}\\ &+  {\zeta^{ab}}{}_{\hat{A}}\wedge\left({M_{abc}}{}^{\hat{A}}\varepsilon^{cdef}e_d\wedge e_e \wedge e_f- F^{\hat{A}} \wedge e_a \wedge e_b\right)\vphantom{\ds\int}\\&- \varepsilon_{abcd}e^a\wedge e^b \wedge e^c \wedge e^d \; \left( Y_{\hat{A}\hat{B}\hat{C}} D^{\hat{A}} D{}^{\hat{B}} D{}^{\hat{C}}+M_{\hat{A}\hat{B}} D^{\hat{A}}D^{\hat{B}} + L_{\hat{A}\hat{B}\hat{C}\hat{D}} D^{\hat{A}} D{}^{\hat{B}} D{}^{\hat{C}}D{}^{\hat{D}}\right)\\&-4\pi i\, l_p^2\,\varepsilon_{abcd}{e^a\wedge e^b \wedge \beta^c D_{\hat{A}}T^{d}{}^{\hat{A}}{}_{\hat{B}}D{}^{\hat{B}}}\,,\vphantom{\ds\int}\\
    \end{aligned}
\end{equation}
where:
\begin{center}
\begin{tabular}{ c c c c }
$B_{\hat{\alpha}}=
  \begin{bmatrix}
    B_{ab} & B_I & B \\ 
  \end{bmatrix}$,
 & $\mathcal{F}^{\hat{\alpha}}=
  \begin{bmatrix}
    R_{ab} & F_I & F \\
  \end{bmatrix}{}^T$, & $D_{\hat{A}}=\begin{bmatrix} & \psi^{\alpha}{}_{\tilde{L}} & \bar{\psi}{}_{\dot{\alpha}}{}_{\tilde{L}} & \psi^{\alpha}{}_R & \bar{\psi}_{\dot{\alpha}}{}_R & \phi_{\tilde{a}}\\ 
  \end{bmatrix}$,
\end{tabular}
\end{center}
\begin{center}
    \begin{tabular}{c c}
       $\mathcal{H}^{\hat{A}}= \begin{bmatrix} \;
       (\overset{\rightarrow}{\nabla} \gamma{}_{\tilde{L}}){}_{\alpha} \; & \; ({\bar{\gamma}{}_{\tilde{L}}}\overset{\leftarrow}{\nabla}){}^{\dot{\alpha}} \; & \; 
       (\overset{\rightarrow}{\nabla} \gamma{}_{\tilde{R}}){}_{\alpha} \; & \;  ({\bar{\gamma}{}_{\tilde{R}}}\overset{\leftarrow}{\nabla}){}^{\dot{\alpha}} \; &\; \D \gamma_{\tilde{a}} \;\;
       \end{bmatrix}{}^T$, & $\gamma_{\hat{A}}=\begin{bmatrix}
       {\gamma}^{\alpha}{}_{\tilde{L}} & {\bar{\gamma}}_{\dot{\alpha}}{}_{\tilde{L}} & {\gamma}^{\alpha}{}_{\tilde{R}} & {\bar{\gamma}}_{\dot{\alpha}}{}_{\tilde{R}} & \gamma_{\tilde{a}}
       \end{bmatrix}$,    \end{tabular}
\end{center}
\begin{center}
\begin{tabular}{c c c}
 $\lambda^{\hat{\alpha}}=
  \begin{bmatrix}
    -\lambda^{ab} & \lambda^I & \lambda \\
  \end{bmatrix}{}^T$, & ${\zeta}^{cd}{}_{\hat{\alpha}}=
  \begin{bmatrix}
   0 & {\zeta^{cd}}_I & {\zeta^{cd}}\\
  \end{bmatrix}$, & ${\zeta^{ab}}_{\hat{A}}=
  \begin{bmatrix}
      \zeta^{ab} & 0 & 0 \\ 
  \end{bmatrix}$,  \\ 
\end{tabular}
\end{center}
\begin{center}
\begin{tabular}{c c}
 $\lambda^{\hat{A}}=
  \begin{bmatrix}
     \lambda{}_{\alpha}{}_L & \bar{\lambda}{}^{\dot{\alpha}}{}_L & \lambda_{\alpha}{}_R & \bar{\lambda}{}^{\dot{\alpha}}{}_R & \lambda^{\tilde{a}} \\
  \end{bmatrix}{}^T$,
& ${M_{cd\hat\alpha}}=
  \begin{bmatrix}
    \varepsilon_{abcd}& {M_{cd}}_I & {M_{cd}}  \\
  \end{bmatrix}$,
\end{tabular}
\end{center}
$$
{M_{abc}}_{\hat{A}}=
  \begin{bmatrix}
\;      \varepsilon_{abcd}\sigma^d{}_{\alpha\dot{\beta}}\bar{\psi}{}^{\dot{\beta}}{}_L \; &\; \varepsilon_{abcd}\bar{\sigma}{}^d{}^{\dot{\alpha}\beta}\psi_\beta{}_L \;&\;  \varepsilon_{abcd}\sigma^d{}_{\alpha\dot{\beta}}\bar{\psi}{}^{\dot{\beta}}{}_R\; &\; \varepsilon_{abcd} \bar{\sigma}{}^d{}^{\dot{\alpha}\beta}\psi{}_{\beta}{}_R \;&\; M_{abc}{}_{\tilde{a}} \;\; \end{bmatrix} .
$$

The matrices $C^{\hat{\alpha}}{}_{\hat{\beta}}$, $C^{\hat{A}}{}_{\hat{B}}$, $M_{\hat{A}\hat{B}}$, $Y_{\hat{A}\hat{B}\hat{C}}$,   $L_{\hat{A}\hat{B}\hat{C}\hat{D}}$ and $T^{d\hat{A}}{}_{\hat{B}}$ are constant matrices, and carry the information about gauge coupling constants, mass of the Higgs field, Yukawa couplings and mixing angles, Higgs self-coupling constant and torsion coupling, respectively.

\section{\label{SecV}Conclusions}

Let us summarize the results of the paper. In section \ref{SecII} we have given a short reminder of the $BF$ theory and described how one can use it to construct the action for general relativity (the well known Plebanski model), and the action for the Yang-Mills theory in flat spacetime, in a novel way. Passing on to higher gauge theory, we have reviewed the formalism of $2$-groups and the corresponding $2BF$ theory, using it again to construct the action for general relativity (a model first described in \cite{MikovicVojinovic2012}), and the unified action of general relativity and Yang-Mills theory, both naturally described using the $2$-group formalism. With this background material in hand, in section \ref{SecIII} we have used the idea of a categorical ladder yet again, generalizing the $2BF$ theory to $3BF$ theory, with the underlying structure of a $3$-group instead of a $2$-group. This has led us to the main insight that the {\em scalar and fermion fields can be specified using a gauge group}, namely the third gauge group, denoted $L$, present in the $2$-crossed module corresponding to a given $3$-group. This has allowed us to single out specific gauge groups corresponding to the Klein-Gordon, Dirac, Weyl and Majorana fields, and to construct the relevant constrained $3BF$ actions that describe all these fields coupled to gravity in the standard way.

The obtained results represent the fundamental building blocks for the construction of the complete Standard Model of elementary particles coupled to Einstein-Cartan gravity as a $3BF$ action with suitable simplicity constraints, as demonstrated in section \ref{SecIV}. In this way, we can complete the first step of the spinfoam quantization programme for the complete theory of gravity and all matter fields, as specified in the Introduction. This is a clear improvement over the ordinary spinfoam models based on an ordinary constrained $BF$ theory.

In addition to this, the gauge group which determines the matter spectrum of the theory is a completely novel structure, not present in the Standard Model. This new gauge group stems from the $3$-group structure of the theory, so it is not surprising that it is invisible in the ordinary formulation of the Standard Model, since the latter does not use any $3$-group structure in an explicit way. In this paper, we have discussed the choices of this group which give rise to all relevant matter fields, and these can simply be directly multiplied to give the group corresponding to the full Standard Model, encoding the quark and lepton families and all other structure of the matter spectrum. However, the true potential of the matter gauge group lies in a possibility of nontrivial unification of matter fields, by choosing it to be something other than the ordinary product of its component groups. For example, instead of choosing $\realni^8(\grasmanovi)$ for the Dirac field, one can try a noncommutative $SU(3)$ group, which also contains $8$ generators, but its noncommutativity requires that the maps $\delta$ and $\{\_\,,\,\_\}$ be nontrivial, in order to satisfy the axioms of a $2$-crossed module. This, in turn, leads to a distinction between $3$-curvature and fake $3$-curvature, which can have consequences for the dynamics of the theory. In this way, by studying nontrivial choices of a $3$-group, one can construct various different $3$-group-unified models of gravity and matter fields, within the context of higher gauge theory. This idea resembles the ordinary grand unification programme within the framework of the standard gauge theory, where one constructs various different models of vector fields by making various choices for the Yang-Mills gauge group. The detailed discussion of these $3$-group unified models is left for future work.

As far as the spinfoam quantization programme is concerned, having completed the step 1 (as outlined in the Introduction), there is a clear possibility to complete the steps 2 and 3 as well. First, the fact that the full action is written completely in terms of differential forms of various degrees, allows us to adapt it to a triangulated spacetime manifold, in the sense of Regge calculus. In particular, all fields and their field strengths present in the $3BF$ action can be naturally associated to the appropriate $d$-dimensional simplices of a $4$-dimensional triangulation, by matching $0$-forms to vertices, $1$-forms to edges, etc. This leads us to the following table:
\begin{center}
\setlength{\tabcolsep}{0pt}
    \label{tab:table4}
\begin{tabular}{|c|c|c|c|c|c|c|} \hline
\ \ $d$\ \ \  & \ \ triangulation$\vphantom{\ds\int}$\ \ \  &\ \  dual triangulation\ \ \  & \ \ form\ \ \   &\ \  fields\ \ \  &\ \ field strengths\ \ \  \\ \hline\hline
$0$ & vertex$\vphantom{\ds\int}$ & $4$-polytope & $0$-form & $\phi$, $\psi_{\tilde\alpha}$, $\bar{\psi}{}^{\tilde\alpha}$ & \\ \hline
$1$ & edge$\vphantom{\ds\int}$ & $3$-polyhedron & $1$-form & \ \  $\omega^{ab}$, $A^I$, $e^a$ \ \ \  &  \\ \hline
$2$ & triangle$\vphantom{\ds\int}$ & face & $2$-form & $\beta^a$, $B^{ab}$ & $R^{ab}$, $F^I$, $T^a$ \\ \hline
$3$ & tetrahedron$\vphantom{\ds\int}$ & edge & $3$-form & $\gamma$, $\gamma_{\tilde\alpha}$, $\bar\gamma{}^{\tilde\alpha}$  & $\cG^a$ \\ \hline
$4$ & $4$-simplex$\vphantom{\ds\int}$ & vertex & \ \  $4$-form\ \ \  & & $\cH$, $\cH_{\tilde\alpha}$, $\bar\cH{}^{\tilde\alpha}$ \\ \hline
\end{tabular}
\end{center}

Once the classical Regge-discretized topological $3BF$ action is constructed, one can attempt to construct a state sum $Z$ which defines the path integral for the theory. The topological nature of the pure $3BF$ action, together with the underlying structure of the $3$-group, should ensure that such a state sum $Z$ is a topological invariant, in the sense that it is triangulation independent. Unfortunately, in order to perform this step precisely, one needs a generalization of the Peter-Weyl and Plancharel theorems to $2$-groups and $3$-groups, a mathematical result that is presently still missing. The purpose of the Peter-Weyl theorem is to provide a decomposition of a function on a group into a sum over the corresponding irreducible representations, which ultimately specifies the appropriate spectrum of labels for the $d$-simplices in the triangulation, fixing the domain of values for the fields living on those $d$-simplices. In the case of $2$-groups and especially $3$-groups, the representation theory has not been developed well enough to allow for such a construction, with a consequence of the missing Peter-Weyl theorem for $2$-groups and $3$-groups. However, until the theorem is proved, we can still try to {\em guess} the appropriate structure of the irreducible representations of the $2$- and $3$-groups, as was done for example in \cite{MikovicVojinovic2012}, leading to the so-called {\em spincube model} of quantum gravity.

Finally, if we remember that for the purpose of physics we are not really interested in a topological theory, but instead in one which contains local propagating degrees of freedom, we are therefore not really engaged in constructing a topological invariant $Z$, but rather a state sum which describes nontrivial dynamics. In particular, we need to impose the simplicity constraints onto the state sum $Z$, which is the step $3$ of the spinfoam quantization programme. In light of that, one of the main motivations and also main results of our paper was to rewrite the action for gravity and matter in a way that explicitly distinguishes the topological sector from the simplicity constraints. Imposing the constraints is therefore straightforward in the context of a $3$-group gauge theory, and completing this step would ultimately lead us to a state sum corresponding to a tentative theory of quantum gravity with matter. This is also a topic for future work.

In the end, let us also mention that aside from the unification and quantization programmes, there is also a plethora of additional studies one can perform with the constrained $3BF$ action, such as the analysis of the Hamiltonian structure of the theory (suitable for a potential canonical quantization programme), the idea of imposing the simplicity constraints using a spontaneous symmetry breaking mechanism, and finally a detailed study of the mathematical structure and properties of the simplicity constraints. This list is of course not conclusive, and there may be many more interesting related topics to study in both physics and mathematics.

\acknowledgments

The authors would like to thank Aleksandar Mikovi\'c, Jeffrey Morton, John Baez, Roger Picken and John Huerta for helpful discussions, comments, and suggestions. This work was supported by the project ON171031 of the Ministry of Education, Science and Technological Development (MPNTR) of the Republic of Serbia, and partially by the bilateral scientific cooperation between Austria and Serbia through the project ``Causality in Quantum Mechanics and Quantum Gravity - 2018-2019'', no. 451-03-02141/2017-09/02, supported by the Federal Ministry of Science, Research and Economy (BMWFW) of the Republic of Austria, and the Ministry of Education, Science and Technological Development (MPNTR) of the Republic of Serbia.

\appendix
\section{\label{ApendiksA}Category theory, $2$-groups and $3$-groups}
\begin{Definition}[Pre-crossed module and crossed module]
A pre-crossed module \\ $(H \stackrel{\del}{\to}G \,, \rhd)$ of groups $G$ and $H$, is given by a group map $\partial : H \to G$,
together with a left action $\rhd$ of $G$ on $H$, by automorphisms, such that for each $h_1\,,h_2 \in H$ and $g \in G$ the following identity hold:
$$g \partial h g^{-1} = \partial (g \rhd h)\,.$$
In a pre-crossed module the {\bf Peiffer commutator} is defined as:
$$
    \langle h_1\,,h_2 \rangle{}_{\mathrm{p}}=h_1h_2h_1^{-1} \partial(h_1) \rhd h_2^{-1}\,.
$$
A pre-crossed module is said to be a {\bf crossed module} if all of its Peiffer commutators are trivial, which is to say that
$$
(\partial h) \rhd h' = hh'h^{-1}\,,
$$
i.e. the {\bf Peiffer identity} is satisfied.
\end{Definition}

\begin{Definition}[$2$-crossed module]
A $2$-crossed module $(L\stackrel{\delta}{\to} H \stackrel{\partial}{\to}G,\,\rhd,\,\{-,\,-\})$ is given by three groups $G$, $H$ and $L$, together with maps $\partial$ and $\delta$ such that:
\begin{displaymath}
    L\stackrel{\delta}{\to} H \stackrel{\partial}{\to}G\,,
\end{displaymath}
where $\partial\delta=1$, an action $\rhd$ of the group $G$ on all three groups, and an $G$-equivariant map called the {\bf Peiffer lifting}:
\begin{displaymath}
\{ - \,,- \} : H\times H \to L\,.
\end{displaymath}
The following identities are satisfied:
\begin{enumerate}
    \item The maps $\partial$ and $\delta$ are $G$-equivariant, i.e. for each $ g \in G$ and $h \in H $:
\begin{equation*}\label{eq:ekvivarijantnost_partial}
    g\rhd \partial (h)=\partial(g\rhd h)\,, \quad \quad \quad g\rhd \delta (l) = \delta(g \rhd l)\,,
\end{equation*}
the action of the group $G$ on the groups $H$ and $L$ is a smooth left action by automorphisms, i.e. for each $g,g_1,g_2 \in G$, $\;h_1,h_2 \in H$, $\;l_1,l_2 \in L$ and $e \in H, L$:
\begin{equation*}\label{eq:ekvivarijantnost_delta}
    g_1 \rhd (g_2 \rhd e) =(g_1 g_2) \rhd e\,, \quad g \rhd (h_1 h_2) =(g \rhd h_1)(g \rhd h_2)\,, \quad g \rhd (l_1 l_2) =(g \rhd l_1)(g \rhd l_2)\,, 
\end{equation*}
and the Peiffer lifting is $G$-equivariant, i.e. for each $h_1,\, h_2 \in H$ and $g \in G$:
$$ g\rhd\{h_1\,,h_2\}=\{ g\rhd h_1, \,g \rhd h_2\}\,;$$
\item the action of the group $G$ on itself is via conjugation, i.e. for each $g\,,g_0 \in G$:
$$ g \rhd g_0 = g\,g_0\,g^{-1}\,;$$
\item In a $2$-crossed module the structure $(L \stackrel{\delta}{\to}H,\,\rhd')$ is a crossed module, with action of the group $H$ on the group $L$ is defined for each $h \in H$ and $l \in L$ as:
$$h \rhd' l = l \, \{\delta(l){}^{-1},\,h\}\,,$$
but $(H \stackrel{\del}{\to}G \,, \rhd)$ may not be one, and the Peiffer identity does not necessary hold. However, when $\partial$ is chosen to be trivial and group $H$ Abelian, the Peiffer identity is satisfied, i.e. for each $h,\,h' \in H$:
$$\delta(h) \rhd h' = h\, h'\, h^{-1}\,;$$
\item $\delta(\{h_1,h_2 \})= \langle h_1\,,h_2\rangle{}_{\mathrm{p}}$, $\qquad \forall h_1,h_2 \in H$, 
\item $[l_1,l_2]=\{\delta(l_1)\,,\delta(l_2) \}$, $\qquad\forall l_1\,, l_2 \in L$. 
Here, the notation $[l,k]=lkl^{-1}k^{-1}$ is used;
\item $\{h_1h_2,h_3 \}=\{h_1,h_2h_3h_2^{-1} \}\partial(h_1)\rhd \{h_2,h_3 \}$, $\qquad\forall h_1,h_2,h_3 \in H$;

\item $\{h_1,h_2h_3 \}= \{h_1,h_2\} \{h_1,h_3\}\{ \langle h_1,h_3 \rangle{}_{\mathrm{p}}^{-1}, \partial(h_1) \rhd h_2\}$,  {$\qquad\forall h_1,h_2,h_3 \in H$};
\item $\{\delta(l),h\}\{h, \delta(l) \}=l(\partial(h) \rhd l^{-1})$, $\qquad \forall h\in H\,, \quad \forall l \in L$.
\end{enumerate}
\end{Definition}
\begin{Definition}[Differential pre-crossed module, differential crossed module]
\  \\ A differential pre-crossed module $(\mathfrak{h} \stackrel{\del}{\to}\mathfrak{g} \,, \rhd)$ of algebras $\mathfrak{g}$ and $\mathfrak{h}$ is given by a Lie algebra map $\partial: \mathfrak{h} \to \mathfrak{g}$ together with an action $\rhd$ of $\mathfrak{g}$ on $\mathfrak{h}$ such that for each $\underline{h} \in \mathfrak{h}$ and $\underline{g} \in \mathfrak{g}$:
$$\partial(\underline{g} \rhd \underline{h})=[\underline{g},\partial(\underline{h})]\,.$$
The action $\rhd$ of $\mathfrak{g}$ on $\mathfrak{h}$ is on left by derivations, i.e. for each $\underline{h}_1, \underline{h}_2 \in \mathfrak{h}$ and each $\underline{g}\in \mathfrak{g}$:
$$ \underline{g} \rhd [\underline{h}_1,\,\underline{h}_2]= [\underline{g} \rhd \underline{h}_1,\,\underline{h}_2] + [\underline{h}_1,\,\underline{g} \rhd \underline{h}_2]\,.$$
In a differential pre-crossed module, the Peiffer commutators are defined for each $\underline{h}{}_1, \underline{h}{}_2 \in \mathfrak{h}$ as:
$$\langle \underline{h}{}_1, \, \underline{h}{}_2 \rangle{}_\mathrm{p}=[\underline{h}{}_1,\underline{h}{}_2]- \partial(\underline{h}{}_1) \rhd \underline{h}{}_2\,.$$
The map $(\underline{h}{}_1, \, \underline{h}{}_2) \in \mathfrak{h} \times \mathfrak{h} \to \langle\underline{h}{}_1, \, \underline{h}{}_2\rangle{}_{\mathrm{p}}\in \mathfrak{h}$ is bilinear $\mathfrak{g}$-equivariant map called the {\bf Peiffer paring}, i.e. all $\underline{h}{}_1 \,, \underline{h}{}_2 \in \mathfrak{h}$ and $\underline{g} \in \mathfrak{g}$ satisfy the following identity:
$$ \underline{g} \rhd \langle \underline{h}{}_1 \,, \underline{h}{}_2 \rangle{}_{\mathrm{p}} = \langle \underline{g} \rhd \underline{h}{}_1 \,, \underline{h}{}_2\rangle + \langle \underline{h}{}_1 \,, \underline{g} \rhd\underline{h}{}_2 \rangle{}_{\mathrm{p}}\,. $$
A differential pre-crossed module is said to be a {\bf differential crossed module} if all of its Peiffer commutators vanish, which is to say that  for each $\underline{h}{}_1, \underline{h}{}_2 \in \mathfrak{h}$:
$$\partial(\underline{h}{}_1)\rhd \underline{h}{}_2=[\underline{h}{}_1,\,\underline{h}{}_2]\,.$$
\end{Definition}
\begin{Definition}[Differential $2$-crossed module] A differential $2$-crossed module is given by a complex of Lie algebras:
\begin{displaymath}
    \mathfrak{l}\stackrel{\delta}{\to} \mathfrak{h} \stackrel{\partial}{\to}\mathfrak{g}\,,
\end{displaymath}
together with left action $\rhd$ of $\mathfrak{g}$ on $\mathfrak{h}$, $\mathfrak{l}$, by derivations, and on itself via adjoint representation, and a $\mathfrak{g}$-equivariant bilinear map called the {\bf Peiffer lifting}:
$$\{-\,,\,-\} : \mathfrak{h} \times \mathfrak{h} \to \mathfrak{l}$$
Fixing the basis in algebra $T_A \in \mathfrak{l}$, $t_a \in \mathfrak{h}$ and $\tau_\alpha \in \mathfrak{g}$:
$$
    [T_A,T_B]={f_{AB}}^C \, T_C\,, \quad \quad [t_a,t_b]={f_{ab}}^c \, t_c\,, \quad \quad  [\tau_\alpha,\tau_\beta]={f_{\alpha\beta}}^\gamma \, \tau_\gamma\,,
$$
one defines the maps $\partial$ and $\delta$ as:
$$
    \partial (t_a)={\partial_a}^\alpha \, \tau_\alpha\,, \quad \quad \quad \delta (T_A)={\delta_A}^a \, t_a\,,
$$
and action of $\mathfrak{g}$ on the generators of $\mathfrak{l}$, $\mathfrak{h}$ and $\mathfrak{g}$ is, respectively:
$$
    \tau_\alpha \rhd T_A={\rhd_{\alpha A}}^B \,T_B\,, \quad \quad \tau_\alpha \rhd t_a={\rhd_{\alpha a}}^b\, t_b\,, \quad \quad \tau_\alpha \rhd \tau_\beta = {\rhd_{\alpha\beta}}^\gamma\, \tau_\gamma \,.
$$
Note that when $\eta$ is $\mathfrak{g}$-valued differential form and $\omega$ is $\mathfrak{l}$, $\mathfrak{h}$ or $\mathfrak{g}$ valued differential form the previous action is defined as:
$$
\eta \rhd \omega = \eta^\alpha \wedge \omega^A \rhd_{\alpha A }{}^B \, T_B\,, \quad \quad \eta \rhd \omega = \eta^\alpha \wedge \omega^a \rhd_{\alpha a }{}^b \, t_b\,, \quad \quad \eta \rhd \omega = \eta^\alpha \wedge \omega^\beta f_{\alpha \beta}{}^\gamma \, \tau_\gamma\,.
$$
The coefficients ${X_{ab}}^A$ are introduced as:
$$
\{t_a,\,t_b\}={X_{ab}}^A T_A\,.
$$
The following identities are satisfied:
\begin{enumerate}
\item In the differential crossed module $(L \stackrel{\delta}{\to}H \,, \rhd')$ the action $\rhd'$ of $\mathfrak{h}$ on $\mathfrak{l}$ is defined for each $\underline{h} \in \mathfrak{h}$ and $\underline{l} \in \mathfrak{l}$ as:
$$\underline{h} \rhd' \underline{l} =-\{\delta(\underline{l}),\,\underline{h}\}\,,$$
or written in the basis where $t_{a} \rhd' T_A = \rhd'{}_{aA}{}^B T_B$ the previous identity becomes:
$${{\rhd'}_{a A}}^B=-{\delta_A}^b {X_{ba}}^B\,;$$
\item The action of $\mathfrak{g}$ on itself is via adjoint representation:
$$ 
    {\rhd_{\alpha\beta}}^\gamma={f_{\alpha\beta}}^\gamma\,;
$$
\item The action of $\mathfrak{g}$ on $\mathfrak{h}$ and $\mathfrak{l}$ is equivariant, i.e. the following identities are satisfied:
$$\partial_a{}^\beta f_{\alpha \beta}{}^\gamma = \rhd_{\alpha a}{}^b \partial_b{}^\gamma\,, \quad \quad \delta_A{}^a\rhd_{\alpha a}{}^b=\rhd_{\alpha A}{}^B \delta_B{}^b\,;$$
\item The Peiffer lifting is $\mathfrak{g}$-equivariant, i.e. for each $\underline{h}_1,\underline{h}_2\in\mathfrak{h}$ and $\underline{g}\in\mathfrak{g}$:
$$\underline{g}\rhd\{\underline{h}_1,\underline{h}_2\}=\{\underline{g}\rhd \underline{h}_1,\underline{h}_2\}+\{\underline{h}_1,\,\underline{g}\rhd \underline{h}_2\}\,,$$
or written in the basis:
$${X_{ab}}^B {\rhd_{\alpha B}}^A={\rhd_{\alpha a}}^c {X_{cb}}^A+{\rhd_{\alpha b}}^c {X_{ac}}^A\,;$$
\item $\delta(\left\{\underline{h}_1,\,\underline{h}_2  \right \})=\left \langle\underline{h}_1,\,\underline{h}_2\right \rangle{}_{\mathrm{p}}\,, \qquad \forall \underline{h}_1,\,\underline{h}_2 \in \mathfrak{h}$, i.e.
$${X_{ab}}^A {\delta_A}^c=f_{ab}{}^c-{\partial_a}^\alpha{\rhd_{\alpha b}}^c\,;$$
\item $[\underline{l}_1,\,\underline{l}_2]=\left\{\delta(\underline{l}_1),\,\delta(\underline{l}_2)\right \}\,, \qquad \forall \underline{l}_1,\,\underline{l}_2 \in \mathfrak{l}$, i.e.
$${f_{AB}}^C={\delta_A}^a{\delta_B}^b {X_{ab}}^C\,;$$
\item $\left\{[\underline{h}_1,\,\underline{h}_2],\,\underline{h}_3\right \}=\partial(\underline{h}_1)\rhd\left\{\underline{h}_2,\,\underline{h}_3\right \}+\left\{\underline{h}_1,\,[\underline{h}_2,\,\underline{h}_3]\right \}-\partial(\underline{h}_2)\rhd\left\{\underline{h}_1,\,\underline{h}_3\right \}-\left\{\underline{h}_2,\,[\underline{h}_1,\,\underline{h}_3]\right \}\,,$ $\forall \underline{h}_1,\,\underline{h}_2,\,\underline{h}_3 \in \mathfrak{h}$, i.e.
$$\left\{[\underline{h}_1,\underline{h}_2],\underline{h}_3\right \}=\{ \partial(\underline{h}_1) \rhd \underline{h}_2,\underline{h}_3\}-\{ \partial(\underline{h}_2) \rhd \underline{h}_1,\underline{h}_3\}-\{\underline{h}_1,\delta\{\underline{h}_2,\underline{h}_3\}\}+\{\underline{h}_2,\delta\{\underline{h}_1,q,\underline{h}_3\}\},$$
$${f_{ab}}^d {X_{dc}}^B={\partial_a}^\alpha {X_{bc}}^A {\rhd_{\alpha A}}^B+{X_{ad}}^B {f_{bc}}^d-{\partial_b}^\alpha{\rhd_{\alpha A}}^B {X_{ac}}^A-{X_{bd}}^B{f_{ac}}^d\,;$$
\item $\left\{\underline{h}_1,\,[\underline{h}_2,\,\underline{h}_3]\right \}=\left\{\delta\left\{\underline{h}_1,\,\underline{h}_2\right \},\underline{h}_3\right \}-\left\{\delta\left\{\underline{h}_1,\,\underline{h}_3\right \},\,\underline{h}_2\right \}\,, \qquad \forall \underline{h}_1,\,\underline{h}_2,\,\underline{h}_3 \in \mathfrak{h}$, i.e.
$${X_{ad}}^A {f_{bc}}^d={X_{ab}}^B{\delta_B}^d {X_{dc}}^A-{X_{ac}}^B{\delta_B}^d {X_{db}}^A\,;$$
\item $\left\{\delta(\underline{l}),\,\underline{h}\right \}+\left\{\underline{h},\,\delta(\underline{l})\right \}=-\partial(\underline{h}) \rhd \underline{l}\,, \qquad \forall \underline{l} \in \mathfrak{l}\,, \quad \forall \underline{h} \in \mathfrak{h}$, i.e.
$${\delta_A}^a {X_{ab}}^B+{\delta_A}^a {X_{ba}}^B=-{\partial_b}^\alpha {\rhd_{\alpha A}}^B\,.$$
\end{enumerate}
\end{Definition}
Note that the property $6.$ implies that either trivial map $\delta$ or the trivial Peiffer lifting imply that $L$ is an Abelian group. Conversely, if $L$ is Abelian, property $6.$ implies that either the map $\delta$ or the Peiffer lifting is trivial, or both.

In the case of an Abelian group $H$ and trivial map $\partial$, among the aforementioned properties the only non-trivial remaining are:
\begin{enumerate}
\item $\delta\{\underline{h}_1,\,\underline{h}_2\}=0\,, \qquad \forall \underline{h}_1\,,\underline{h}_2 \in \mathfrak{h}\,;$
\item $[\underline{l}_1,\,\underline{l}_2]=\{\delta(\underline{l}_1),\,\delta(\underline{l}_2)\}\,, \qquad \forall\underline{l}_1\,,\underline{l}_2 \in \mathfrak{l}\,;$
\item $\{\delta(\underline{l}),\,\underline{h}\}=-\{\underline{h},\,\delta(\underline{l})\}\,, \qquad \forall \underline{h} \in \mathfrak{h}\,, \quad \forall\underline{l} \in \mathfrak{l}\,.$
\end{enumerate}
A reader intrested in more details about $3$-groups is referred to \cite{Wang2014}.

\section{\label{ApendiksB}The construction of gauge-invariant actions for $3BF$ theory}
Symmetric bilinear invariant nondegenerate forms are defined as:
$$
    \langle T_A\,, T_B \rangle{}_\mathfrak{l}=g{}_{AB}\,, \quad \quad \langle t_a\,, t_b \rangle{}_\mathfrak{h}=g{}_{ab}\,, \quad \quad \langle \tau_\alpha\,,\tau_\beta \rangle{}_\mathfrak{g}=g{}_{\alpha\beta}\,.
$$
They satisfy the following properties:
\begin{itemize}
    \item $\langle \_\,,\_ \rangle{}_\mathfrak{g}$ is $G$-invariant:
    $$
        \langle g \tau_\alpha g^{-1}\,, g \tau_\beta g^{-1}\rangle{}_\mathfrak{g}=\langle\tau_\alpha\,,\tau_\beta\rangle{}_\mathfrak{g}\,, \quad \forall g \in G\,;
    $$
    \item $\langle \_\,,\_ \rangle{}_\mathfrak{h}$ is $G$-invariant:
    $$
        \langle g \rhd  t_a\,, g \rhd t_b\rangle{}_\mathfrak{h}=\langle t_a\,,t_b\rangle{}_\mathfrak{h}\,, \quad \forall g \in G\,,
    $$
    and, when $(H \stackrel{\del}{\to}G\,, \rhd)$ is a crossed module, consequently $H$-invariant:
    $$
        \langle ht_ah^{-1}\,,ht_b h^{-1}\rangle{}_\mathfrak{h}=\langle\partial(h)\rhd t_a\,, \partial(h)\rhd t_b\rangle{}_\mathfrak{h}=\langle t_a\,,t_b\rangle{}_\mathfrak{h}\,, \quad  \forall h \in H\,;
    $$
    \item $\langle \_\,,\_ \rangle{}_\mathfrak{l}$ is $G$-invariant:
    $$
        \langle g \rhd  T_A\,, g \rhd T_B\rangle{}_\mathfrak{l}=\langle T_A\,,T_B\rangle{}_\mathfrak{l}\,,  \quad \forall g \in G\,,
    $$
    and in the case when the Peiffer lifting or the map $\delta$ is trivial consequently $H$-invariant:
    $$
         \langle h \rhd'  T_A\,, h \rhd' T_B\rangle{}_\mathfrak{l}=\langle T_A - \{\delta(T_A),\,h\}\,,T_B - \{\delta(T_B),\,h\}\rangle{}_\mathfrak{l}=\langle T_A\,,T_B\rangle{}_\mathfrak{l}\,, \quad \forall h \in H\,. 
    $$
    From the $H$-invariance of $\langle \_\,,\_ \rangle{}_\mathfrak{l}$ and properties of a crossed module $(L \stackrel{\delta}{\to}H \,, \rhd')$ follows $L$-invariance:
    $$
        \langle lT_Al^{-1}\,,lT_Bl^{-1}\rangle{}_\mathfrak{l}=\langle \delta (l)\rhd' T_A\,,\delta(l)\rhd'T_B\rangle{}_\mathfrak{l}=\langle T_A\,,T_B\rangle{}_\mathfrak{l}\,, \quad \forall l \in L\,.
    $$
\end{itemize}
From the invariance of the bilinear forms follows the existence of gauge-invariant topological $3BF$ action of the form:
\begin{equation}
S_{3BF} =\int_{\mathcal{M}_4} \langle B \wedge \cal F \rangle_{\mathfrak{g}} +  \langle C \wedge \cal G \rangle_{\mathfrak{h}} + \langle D \wedge \cal H \rangle_{\mathfrak{l}} \,,
\end{equation}
where $B \in \cA^2(\cM_4\,,\mathfrak{g})$, $C \in \cA^1(\cM_4\,,\mathfrak{h})$ and $D \in \cA^0(\cM_4\,,\mathfrak{l})$ are Lagrange multipliers, and ${\cal F} \in \cA^2(\cM_4\,,\mathfrak{g})$, ${\cal G} \in \cA^3(\cM_4\,,\mathfrak{h})$ and ${\cal H} \in \cA^4(\cM_4\,,\mathfrak{l})$ are curvatures defined as in (\ref{eq:3krivine}). Written in the basis:
\begin{equation*}
\begin{array}{c}
\ds {\cal F}=\frac{1}{2}{{\cal F}^\alpha{}_{\mu\nu}} \tau_\alpha \D x^\mu \wedge \D x^\nu\,, \qquad {\cal G}=\frac{1}{3!}{{\cal G}^a{}_{\mu \nu \rho}} t_a \D x^\mu \wedge \D x^\nu \wedge \D x^\rho\,, \vphantom{\ds\int} \\
\ds {\cal H}=\frac{1}{4!}{\cal H }^A{}_{\mu\nu\rho\sigma} T_A\D x^\mu \wedge \D x^\nu \wedge \D x^\rho \wedge \D x^\sigma\,, \vphantom{\ds\int} \\
\end{array}
\end{equation*}
the coefficients are:
\begin{gather*} 
\begin{aligned}
{{\cal F}^\alpha{}_{\mu \nu}}=&\partial_{\mu} {\alpha^\alpha{}_{\nu}}-\partial_{\nu}{\alpha^\alpha{}_{\mu}} + {f_{\beta\gamma}}^\alpha {\alpha}^\beta{}_{\mu}{\alpha}^\gamma{}_{\nu} -\beta^a{}_{\mu\nu}\partial_a{}^\alpha\,,\\
{{\cal G}^a{}_{\mu \nu\rho}}=&\partial_{\mu}\beta^a{}_{\nu \rho}+\partial_{\nu}\beta^a{}_{\rho \mu}+\partial_{\rho}{\beta^a{}_{\mu \nu}} \\
&+{\alpha}^\alpha{}_{\mu}\beta^b{}_{\nu \rho}{\rhd_{\alpha b}}^a+{\alpha}^\alpha{}_{\nu}\beta^b{}_{\rho \mu}{\rhd_{\alpha b}}^a+{\alpha}^\alpha{}_{\rho}\beta^b{}_{\mu \nu}{\rhd_{\alpha b}}^a-\gamma^A{}_{\mu\nu\rho}{}\delta_A{}^a\,,\\
{{\mathcal{H}}^A{}_{\mu \nu \rho \sigma}}=&\partial_{\mu} {\gamma^A{}_{\nu \rho \sigma}}-\partial_{\nu} {\gamma^A{}_{\rho \sigma\mu}}+\partial_{\rho} {\gamma^A{}_{ \sigma\mu\nu}}-\partial_{\sigma} {\gamma^A{}_{\mu\nu \rho}}\\&+2\beta^a{}_{\mu \nu} \beta^b{}_{\rho\sigma} X_{\{ab\}}{}^A-2\beta^a{}_{\mu\rho}\beta^b{}_{\nu\sigma} X_{\{ab\}}{}^A+2\beta^a{}_{\mu\sigma}\beta^b{}_{\nu\rho} X_{\{ab\}}{}^A\\&+{\alpha}^\alpha{}_{\mu}{\gamma^B{}_{\nu\rho\sigma}}{\rhd_{\alpha B}}^A-{\alpha}^\alpha{}_{\nu}{\gamma^B{}_{\rho\sigma\mu}}{\rhd_{\alpha B}}^A+{\alpha}^\alpha{}_{\rho}{\gamma^B{}_{\sigma\mu\nu}}{\rhd_{\alpha B}}^A-{\alpha}^\alpha{}_{\sigma}{\gamma^B{}_{\mu\nu\rho}}{\rhd_{\alpha B}}^A\,.    
\end{aligned}
\end{gather*}

Note that the wedge product $A \wedge B$ when $A$ is a $0$-form and $B$ is a $p$-form is defined as $A \wedge B = \frac{1}{p!}A B_{\mu_1 \dots \mu_p} \D x^{\mu_1} \wedge \dots \wedge x^{\mu_p}$.

Given $G$-invariant symmetric non-degenerate bilinear forms in $\mathfrak{g}$ and $\mathfrak{h}$, one can define a bilinear antisymmetric map ${\cal T}: \mathfrak{h} \times \mathfrak{h} \to \mathfrak{g}$ by the rule:
$$
    \langle {\cal T}(\underline{h}_1,\,\underline{h}_2)\,, \underline{g}\rangle{}_\mathfrak{g}=-\langle \underline{h}_1,\,\underline{g}\rhd \underline{h}_2\rangle{}_\mathfrak{h}, \quad \quad \forall \underline{h}_1,\,\underline{h}_2 \in \mathfrak{h}\,, \quad \forall \underline{g} \in \mathfrak{g}\,.
$$
See \cite{FariaMartinsMikovic2011} for more properties and the construction of $2BF$ invariant topological action using this map. 
To define $3BF$ invariant topological action one has to first define a bilinear antisymmetric map ${\cal S} : \mathfrak{l} \times \mathfrak{l} \to \mathfrak{g}$ by the rule:
$$
    \langle {\cal S} (\underline{l}_1,\,\underline{l}_2),\, \underline{g}\rangle{}_{\mathfrak{g}}=-\langle \underline{l}_1,\,\underline{g}\rhd \underline{l}_2\rangle{}_{\mathfrak{l}}\,, \quad \quad \forall \underline{l}_1, \forall \underline{l}_2 \in \mathfrak{l}\,,\quad \forall \underline{g} \in \mathfrak{g}\,.
$$
Note that $\langle\_\,,\_\rangle{}_\mathfrak{g}$ is non-degenerate and 
$$
\langle  \underline{l}_1,\, \underline{g} \rhd \underline{l}_2 \rangle{}_\mathfrak{l}=- \langle \underline{g} \rhd \underline{l}_1,\, \underline{l}_2\rangle_\mathfrak{l} =- \langle \underline{l}_2,\,\underline{g} \rhd \underline{l}_1\rangle_\mathfrak{l} \,, \quad \quad \forall \underline{g} \in \mathfrak{g}, \quad \forall \underline{l}_1, \, \underline{l}_2 \in \mathfrak{l}\,.$$
Morever, given $g \in G$ and $\underline{l}_1,\, \underline{l}_2 \in \mathfrak{l}$ one has:
$${\cal S}(g \rhd \underline{l}_1,\,g \rhd \underline{l}_2)=g\,{\cal S}(\underline{l}_1,\,\underline{l}_2)\,g^{-1}\,,$$
since for each $\underline{g} \in \mathfrak{g}$ and $\underline{l}_1,\,\underline{l}_2 \in \mathfrak{l}$: 
\begin{equation*}
\begin{aligned}
\langle \underline{g},\,g^{-1} {\cal S}( g \rhd \underline{l}_1\,,g \rhd \underline{l}_2) g \rangle_\mathfrak{g}&=\langle g\underline{g}g^{-1},\,{\cal S}(g \rhd \underline{l}_1,\,g \rhd \underline{l}_2)\rangle_{\mathfrak{g}}\\&=-\langle (g\,\underline{g}\,g^{-1}) \rhd g \rhd  \underline{l}_1,\, g \rhd \underline{l}_2\rangle_{\mathfrak{l}}\\
&=-\langle \underline{g} \rhd \underline{l}_1\,,\underline{l}_2 \rangle_{\mathfrak{l}}=\langle \underline{g}\,, {\cal S}(\underline{l}_1,\,\underline{l}_2)\rangle_\mathfrak{g} \,,   
\end{aligned}
\end{equation*}
where the following mixed relation has been used:
\begin{equation}
    g\rhd(\underline{g} \rhd \underline{l})=(g\,\underline{g}\,g^{-1}) \rhd g \rhd \underline{l}\,.\label{eq:mixed}
\end{equation}
We thus have the following identity:
$${\cal S}(\underline{g} \rhd \underline{l}_1,\,\underline{l}_2)+{\cal S}(\underline{l}_1,\,\underline{g} \rhd \underline{l}_2)=[\underline{g},\,{\cal S}(\underline{l}_1,\,\underline{l}_2)]\,.$$
As far as the bilinear antisymmetric map ${\cal S} : l \times l \to g$, one can write it in the basis:
$${\cal S}(T_A,T_B)={\cal S}{}_{AB}{}^\alpha \tau_\alpha\,,$$
so that the defining relation for ${\cal S}$ becomes the relation:
$${\cal S}_{AB}{}^\alpha g_{\alpha\beta}=-\rhd_{\alpha [B}{}^C g_{A]C}\,.$$
Given two $\mathfrak{l}$-valued forms $\eta$ and $\omega$, one can define a $\mathfrak{g}$-valued form:
$$\omega \wedge^{\cal S} \eta = \omega^A \wedge \eta^B {\cal S}{}_{AB}{}^\alpha \tau_\alpha\,.$$
Now one can define the transformations of the Lagrange multipliers under $L$-gauge  transformations (\ref{eq:LLg}).

Further, to define the transformations of the Lagrange multipliers under $H$-gauge transformations one needs to define the bilinear map ${\cal X}_1: \mathfrak{l} \times \mathfrak{h} \to \mathfrak{h}$ by the rule:
$$\langle {\cal X}_1(\underline{l},\,\underline{h}_1),\,\underline{h}_2 \rangle{}_{\mathfrak{h}}=-\langle \underline{l},\,\{\underline{h}_1,\,\underline{h}_2\}\rangle{}_{\mathfrak{l}}\,,\quad \quad \forall \underline{h}_1,\,\underline{h}_2 \in \mathfrak{h}\,,\quad \forall \underline{l}\in\mathfrak{l}\,, $$
and bilinear map ${\cal X}_2:\mathfrak{l} \times\mathfrak{h} \to\mathfrak{h} $ by the rule:
$$\langle {\cal X}_2(\underline{l},\,\underline{h}_2),\,\underline{h}_1 \rangle{}_{\mathfrak{h}}=-\langle \underline{l},\,\{\underline{h}_1,\,\underline{h}_2\}\rangle{}_{\mathfrak{l}}\,,\quad \quad \forall \underline{h}_1,\,\underline{h}_2 \in \mathfrak{h}\,,\quad \forall \underline{l}\in\mathfrak{l}\,. $$
As far as the bilinear maps ${\cal X}_1$ and ${\cal X}_2$ one can define the coefficients in the basis as:
$${\cal X}_1(T_A,t_a)={\cal X}_1{}_{Aa}{}^b \,t_b\,, \quad \quad {\cal X}_2(T_A,t_a)={\cal X}_2{}_{Aa}{}^b \,t_b\,.$$
When written in the basis the defining relations for the maps ${\cal X}_1$ and ${\cal X}_2$ become:
$$ {\cal X}_1{}_{Ab}{}^cg_{ac}=-X_{ba}{}^Bg_{AB}\,,\quad \quad \quad {\cal X}_2{}_{Ab}{}^cg_{ac}=-X_{ab}{}^Bg_{AB}\,.$$
Given $\mathfrak{l}$-valued differential form $\omega$ and $\mathfrak{h}$-valued differential form $\eta$, one defines a $\mathfrak{h}$-valued form as:
$$\omega\wedge^{{\cal X}_1}\eta=\omega^A\wedge\eta^a{{\cal X}_1}_{Aa}{}^{b}t_b\,, \quad \quad \omega\wedge^{{\cal X}_2}\eta=\omega^A\wedge\eta^a{{\cal X}_2}_{Aa}{}^{b}t_b\,. $$
Given any $g\in G$, $\underline{l} \in \mathfrak{l}$ and $\underline{h} \in \mathfrak{h}$ one has:
$${\cal X}_1(g\rhd \underline{l},\,g^{-1}\rhd \underline{h})=g\rhd {\cal X}_1(\underline{l},\,\underline{h})\,,\quad\quad\quad {\cal X}_2(g\rhd \underline{l},\,g\rhd \underline{h})=g{}^{-1}\rhd {\cal X}_2(\underline{l},\,\underline{h})\,, $$
since for each $\underline{h}_1,\underline{h_2} \in \mathfrak{h}$ and $\underline{l} \in \mathfrak{l}$:
\begin{equation*}
\begin{aligned}
\langle \underline{h}_2,\,g^{-1}\rhd{\cal X}_1(g\rhd \underline{l},\,g\rhd \underline{h}_1)\rangle{}_{\mathfrak{h}}&=\langle g \rhd \underline{h}_2,\,{\cal X}_1(g\rhd \underline{l},\,g\rhd \underline{h}_1)\rangle{}_{\mathfrak{h}}=\langle g\rhd \underline{l},\,\{g\rhd \underline{h}_1,\,g\rhd \underline{h}_2\}\rangle{}_{\mathfrak{l}}\\ \langle g\rhd \underline{l},\,g\rhd\{ \underline{h}_1,\, \underline{h}_2\}\rangle{}_{\mathfrak{l}}&=\langle \underline{l},\,\{ \underline{h}_1,\,\underline{h}_2\}\rangle{}_{\mathfrak{l}}=\langle \underline{h}_2,\,{\cal X}_1(l,\,\underline{h}_1)\rangle{}_{\mathfrak{h}}\,,
\end{aligned}
\end{equation*}
and similarly for $ {\cal X}{}_2$.
Finaly, one needs to define a trilinear map ${\cal D}: \mathfrak{h} \times \mathfrak{h} \times \mathfrak{l} \to \mathfrak{g} $ by the rule:
$$\langle {\cal D}(\underline{h}_1,\,\underline{h}_2,\,\underline{l}),\,\underline{g}\rangle{}_{\mathfrak{g}}=-\langle \underline{l},\,\{\underline{g}\rhd\underline{h}_1,\,\underline{h}_2\}\rangle{}_{\mathfrak{l}}\,,\quad \quad \forall \underline{h}_1,\,\underline{h}_2 \in \mathfrak{h}\,, \quad \forall \underline{l} \in \mathfrak{l},\,\quad \forall \underline{g}\in \mathfrak{g}\,, $$
One can define the coefficients of the trilinear map as:
$${\cal D}(t_a,\,t_b,\,T_A)= {\cal D}_{abA}{}^\alpha \tau_\alpha\,, $$
and the defining relation for the map ${\cal D}$ expressed in terms of coefficients becomes:
$${\cal D}{}_{abA}{}^\beta g_{\alpha\beta}=-\rhd_{\alpha a}{}^c X_{cb}{}^B g_{AB}\,.$$
Given two $\mathfrak{h}$-valued forms $\omega$ and $\eta$, and $\mathfrak{l}$-valued form $\xi$, the $g$-valued form is given by the formula:
$$\omega\wedge^{\cal D}\eta\wedge^{\cal D}\xi=\omega^a\wedge\eta^b\wedge\xi^A {\cal D}{}_{abA}{}^\beta \tau_\beta\,.$$
The following compatibility relation between the maps ${{\cal X}_1}$ and ${\cal D}$ hold:
\begin{equation}\label{eq:comp}
    \langle {\cal D}(\underline{h}_1,\,\underline{h}_2,\,\underline{l}),\,\underline{g}\rangle{}_{\mathfrak{g}}=\langle {\cal X}_1(\underline{l},\,\underline{g}\rhd\underline{h}_1),\,\underline{h}_2 \rangle{}_{\mathfrak{h}}\,,\quad \quad \forall \underline{h}_1,\,\underline{h}_2 \in \mathfrak{h}\,, \quad \forall \underline{l} \in \mathfrak{l},\,\quad \forall \underline{g}\in \mathfrak{g}\,, 
\end{equation}
which one can prove valid from the defining relations in terms of the coefficients.
One can demonstrate that for each $ \underline{h}_1,\,\underline{h}_2 \in \mathfrak{h}$, $\underline{l} \in \mathfrak{l}$ and $g\in G$:
$${\cal D}(g\rhd\underline{h}_1,\,g\rhd\underline{h}_2,\,g\rhd\underline{l})=g\,{\cal D}(\underline{h}_1,\,\underline{h}_2,\,\underline{l})\,g^{-1}\,,$$
since for each $ \underline{h}_1,\,\underline{h}_2 \in \mathfrak{h}$, $\underline{l} \in \mathfrak{l}$, $\underline{g}\in \mathfrak{g}$ and $g \in G$:
\begin{equation*}
\begin{aligned}
    \langle g^{-1}{\cal D}(g\rhd\underline{h}_1,\,g\rhd\underline{h}_2,\,g\rhd\underline{l})g,\,\underline{g}\rangle{}_{\mathfrak{g}}&=\langle {\cal D}(g\rhd\underline{h}_1,\,g\rhd\underline{h}_2,\,g\rhd\underline{l}),\,g\underline{g}g^{-1}\rangle{}_{\mathfrak{g}}\\&=\langle {\cal X}_1(g\rhd\underline{l},\,g\underline{g}g^{-1}\rhd g\rhd\underline{h}_1),\,g\rhd\underline{h}_2 \rangle{}_{\mathfrak{h}}\\&=\langle {\cal X}_1(g\rhd\underline{l},\,g\rhd \underline{g}\rhd\underline{h}_1),\,g\rhd\underline{h}_2 \rangle{}_{\mathfrak{h}}\\&=\langle g\rhd{\cal X}_1(\underline{l},\, \underline{g}\rhd\underline{h}_1),\,g\rhd\underline{h}_2 \rangle{}_{\mathfrak{h}}\\&=\langle {\cal X}_1(\underline{l},\, \underline{g}\rhd\underline{h}_1),\,\underline{h}_2 \rangle{}_{\mathfrak{h}}\\ &=\langle{\cal D}(\underline{h}_1,\,\underline{h}_2,\,\underline{l})\,,\underline{g}\rangle{}_{\mathfrak{g}}\,,
\end{aligned}
\end{equation*}
where the relation (\ref{eq:mixed}) and the compatibility relation (\ref{eq:comp}) were used.
We thus have for each $\underline{h}_1,\,\underline{h}_2\in \mathfrak{h}$, $\underline{l} \in \mathfrak{l}$ and $\underline{g} \in \mathfrak{g}$ the following identity:
$${\cal D}(\underline{g}\rhd\underline{h}_1,\,\underline{h}_2,\,\underline{l})+{\cal D}(\underline{h}_1,\,\underline{g}\rhd\underline{h}_2,\,\underline{l})+{\cal D}(\underline{h}_1,\,\underline{h}_2,\,\underline{g}\rhd\underline{l})=[\underline{g},\,{\cal D}(\underline{h}_1,\,\underline{h}_2,\,\underline{l})]\,. $$
Now one can define the transformations of the Lagrange multipliers under $H$-gauge transformations as in (\ref{eq:LHg}).

\section{\label{ApendiksC}The equations of motion for the Weyl and Majorana fields}

The action for the Weyl spinor field coupled to gravity is given by (\ref{eq:Weyl}). The variation of this action with respect to the variables $B_{ab}$, $\lambda^{ab}$, $\gamma{}_\alpha$, $\bar{\gamma}{}^{\dot{\alpha}}$, $\lambda{}_\alpha$, $\bar{\lambda}{}{}^{\dot{\alpha}}$, $\psi_\alpha$, $\bar{\psi}{}^{\dot{\alpha}}$, $e^a$, $\beta^a$ and $\omega^{ab}$ one obtains the complete set of equations of motion, as follows:
\begin{gather*}
\label{eq:f00}
R^{ab}-\lambda^{ab}=0\,,\vphantom{\ds\int}\\
\label{eq:f000}
B_{ab}-\frac{1}{16\pi l_p^2}\varepsilon_{abcd} e^c \wedge e^d=0\,,\vphantom{\ds\int}\\
\label{eq:f02}
\nabla\psi_\alpha+\lambda{}_\alpha=0\,,\vphantom{\ds\int}\\
\label{eq:f04}
\nabla \bar{\psi}{}^{\dot{\alpha}}+ \bar{\lambda}{}^{\dot{\alpha}}=0\,,\vphantom{\ds\int}\\
\label{eq:f05}
-\gamma{}_\alpha+\frac{i}{6}\varepsilon_{abcd}e^a\wedge e^b \wedge e^c \sigma^d{}_{\alpha\dot{\beta}}\bar{\psi}{}^{\dot{\beta}}=0\,,\vphantom{\ds\int}\\
\label{eq:f06}
-\bar{\gamma}{}^{\dot{\alpha}}+\frac{i}{6}\varepsilon_{abcd}e^a\wedge e^b \wedge e^c \bar{\sigma}{}^d{}^{\dot{\alpha}\beta}\psi_\beta=0\,,\vphantom{\ds\int}\\
\label{eq:f010}
\nabla\gamma{}_\alpha-\frac{i}{6}\varepsilon_{abcd}e^a \wedge e^b \wedge e^c \sigma{}^d{}_{\alpha\dot{\beta}}\bar{\lambda}{}^{\dot{\beta}}=0\,,\vphantom{\ds\int}\\
\label{eq:f012}
\nabla\bar{\gamma}{}^{\dot{\alpha}}-\frac{i}{6}\varepsilon_{abcd}e^a \wedge e^b \wedge e^c \bar{\sigma}{}^d{}^{\dot{\alpha}\beta}\lambda{}_\beta=0\,,\vphantom{\ds\int}\\
\label{eq:f013}
\begin{aligned}
\nabla\beta_a+\frac{1}{8\pi l_p^2}\varepsilon_{abcd} & \lambda^{bc} \wedge e^d+\frac{i}{2} \varepsilon_{abcd} e^b \wedge e^c \wedge (\bar{\lambda}{}_{\dot{\alpha}}\bar{\sigma}{}^d{}^{\dot{\alpha}\beta}\psi_\beta+\lambda{}^\alpha\sigma^d{}_{\alpha\dot{\beta}}\bar{\psi}{}^{\dot{\beta}}) \\
&-8\pi i l_p^2\varepsilon_{abcd}e^b\beta^c\big( \psi^\alpha  (\sigma^d)_{\alpha\dot{\beta}}\bar{\psi}{}^{\dot{\beta}} \big)=0\,,\vphantom{\ds\int}
\end{aligned}\\
\label{eq_f014}
\nabla e_a-4\pi l_p^2 \varepsilon_{abcd}e^b\wedge e^c \wedge (\bar{\psi}{}_{\dot{\alpha}}\bar{\sigma}{}^d{}^{\dot{\alpha}\beta}\psi_\beta)=0\,,\vphantom{\ds\int}\\
\label{eq:f015}
\nabla B_{ab}-e_{[a}\wedge\beta_{b]}-\frac{1}{2}\gamma\sigma{}^{ab}{}_\alpha{}^\beta\psi_\beta-\frac{1}{2}\bar{\gamma}{}_{\dot{\alpha}}\bar{\sigma}{}^{ab}{}^{\dot{\alpha}}{}_{\dot{\beta}}\bar{\psi}{}^{\dot{\beta}}=0\,.\vphantom{\ds\int}
\end{gather*}
In the case of the Majorana field, one adds the mass term (\ref{eq:MajoranaMassTerm}) to the action (\ref{eq:Weyl}). Then, the variation of the action with respect to $B_{ab}$, $\psi^{ab}$, $\gamma^\alpha $, $\bar{\gamma}_{\dot{\alpha}}$, $\lambda_\alpha $, $\bar{\lambda}{}^{\dot{\alpha}} $, $\psi_\alpha $, $\bar{\psi}{}^{\dot{\alpha}}_I$, $e^a$, $\beta^a$ and $\omega_{ab}$ gives the equations of motion for the Majorana case, as follows:
\begin{gather*}\label{eq:m0}
R^{ab}-\lambda^{ab}=0\,,\vphantom{\ds\int}\\\label{eq:m1}
B_{ab}-\frac{1}{16\pi l_p^2}\varepsilon_{abcd} e^c \wedge e^d=0\,,\vphantom{\ds\int}\\ \label{eq:m2}
-\nabla \psi_\alpha  + \lambda_\alpha=0\,,\vphantom{\ds\int}\\ \label{eq:m3}
-\nabla\bar{\psi}^{\dot{\alpha}} +\lambda^{\dot{\alpha}} =0\,,\vphantom{\ds\int}\\  \label{eq:m4}
\gamma^\alpha -\frac{i}{6}\varepsilon_{abcd}e^a\wedge e^b\wedge e^c \bar{\psi}_{\dot{\beta}}  (\bar{\sigma}{}^d){}^{\dot{\beta}\alpha}=0\,,\vphantom{\ds\int}\\ \label{eq:m5}
    \bar{\gamma}{}_{\dot{\alpha}} -\frac{i}{6}\varepsilon_{abcd}e^a\wedge e^b \wedge e^c \psi^\beta (\sigma^d){}_{\beta\dot{\alpha}}=0\,,\vphantom{\ds\int}\\ \label{eq:m6}
\begin{aligned}
    \nabla\gamma^\alpha+\frac{i}{6}\varepsilon_{abcd} \lambda^{\dot{\beta}} & \wedge e^a \wedge e^b \wedge e^c (\sigma^d){}^\alpha{}_{\dot{\beta}}-\frac{1}{6}\ m\varepsilon_{abcd}e^a \wedge e^b \wedge e^c \wedge e^d \psi^\alpha\\&-4i\pi l_p^2 \varepsilon_{abcd}e^a\wedge e^b \wedge \beta^c \bar{\psi}{}_{\dot{\beta}} (\bar{\sigma}{}^d){}^{\dot{\beta}\alpha}=0\,,\vphantom{\ds\int}
\end{aligned}\\ \label{eq:m7}
\begin{aligned}
    \nabla {\bar{\gamma}}{}_{\dot{\alpha}} +\frac{i}{6}\varepsilon_{abcd}\lambda_\beta & \wedge e^a \wedge e^b \wedge e^c (\bar{\sigma}{}^d){}_{\dot{\alpha}}{}^\beta-\frac{1}{6} m\varepsilon_{abcd}e^a \wedge e^b \wedge e^c \wedge e^d \psi_{\dot{\alpha}}\\&-4i\pi l_p^2\varepsilon_{abcd}e^a\wedge e^b \wedge \beta^c {\psi}{}^{{\beta}} (\sigma^d){}_{\beta\dot{\alpha}}=0\,,\vphantom{\ds\int}
\end{aligned}\\ \label{eq:m8}
    \begin{aligned}
    \nabla \beta^a&+\frac{1}{8\pi l_p^2}\varepsilon_{abcd}\lambda^{bc}\wedge e^d+ \frac{i}{2}\varepsilon_{abcd}\lambda_\alpha  \wedge e^b \wedge e^c \bar{\psi}{}_{\dot{\beta}}  (\bar{\sigma}{}^d){}^{\dot{\beta}\alpha}+ \frac{i}{2}\varepsilon_{abcd}\lambda^{\dot{\alpha}}  \wedge e^b \wedge e^c \psi^\beta (\sigma^d){}_{\beta\dot{\alpha}}\vphantom{\ds\int}\\&-\frac{1}{3}m\varepsilon_{abcd}e^b \wedge e^c \wedge e^d ( \psi^\alpha  \psi_\alpha  +\bar{\psi}_{\dot{\alpha}} \bar{\psi}^{\dot{\alpha}} ) -8\pi i l_p^2\varepsilon_{abcd}e^b\beta^c\big( \psi^\alpha  (\sigma^d)_{\alpha\dot{\beta}}\bar{\psi}{}^{\dot{\beta}} \big)=0\,,\vphantom{\ds\int}\end{aligned}\\ \label{eq:m9}
    \nabla e_a-4i\pi l_p^2\varepsilon_{abcd}e^b\wedge e^c \big( \psi^\alpha  (\sigma^d)_{\alpha\dot{\beta}}\bar{\psi}{}^{\dot{\beta}} \big) =0\,,\vphantom{\ds\int}\\ \label{eq:m10}
    \nabla B_{ab}- e_{[a}\wedge \beta_{b]}-\frac{1}{2}\psi^\alpha(\sigma{}^{ab}){}_\alpha{}^\beta \gamma{}_\beta - \frac{1}{2}\bar{\psi}{}_{\dot{\alpha}} (\bar{\sigma}{}^{ab})^{\dot{\alpha}}{}_{\dot{\beta}}\bar{\gamma}{}^{\dot{\beta}} =0\,.\vphantom{\ds\int}
\end{gather*}

\end{document}